\documentclass[a4paper,11pt,reqno]{article}

\usepackage{a4wide}
\usepackage{amsmath,amsfonts,amssymb}
\usepackage[english]{babel}
\usepackage{soul}
\usepackage{nicefrac}
\usepackage[mathscr]{euscript}
\usepackage{setspace}
\usepackage[sort&compress,merge,numbers]{natbib}
\usepackage{comment}
\usepackage{datetime}

\usepackage{mathrsfs}
\usepackage[T1]{fontenc}
\usepackage{mathpazo}
\usepackage{epsfig}

\usepackage[breaklinks=true]{hyperref}

\numberwithin{equation}{section}

\hypersetup{
			colorlinks=true,
			urlcolor=blue,
			citecolor=magenta,
			linkcolor=blue,
     }

\let\oldsqrt\sqrt
\def\sqrt{\mathpalette\DHLhksqrt}

\def\DHLhksqrt#1#2{%
\setbox0=\hbox{$#1\oldsqrt{#2\,}$}\dimen0=\ht0
\advance\dimen0-0.2\ht0
\setbox2=\hbox{\vrule height\ht0 depth -\dimen0}%
{\box0\lower0.4pt\box2}}

\def\beq{\begin{equation}}
\def\eeq{\end{equation}}

\def\ov{\overline}

\author{
  \begin{minipage}{.97\linewidth}
    \vspace{1cm}
       \begin{center}
      \begin{small}
        \textbf{Ignatios Antoniadis},$^{1,2}$
      \textbf{Jean-Pierre Derendinger},$^2$\\
     \textbf{P. Marios Petropoulos}$^{3,1}$ and 
      \textbf{Konstantinos Siampos}$^2$
              \end{small}
    \end{center}
    \vspace{0.5cm}
    \hspace{2.4cm}\begin{minipage}{.7\linewidth}
\begin{center}     {\it \begin{footnotesize}
\hbox{\kern-3.cm\vbox{\vskip0cm
 \begin{itemize}
               \item[$^1$] Laboratoire de Physique Th\'eorique et Hautes Energies,\\ 
        Sorbonne Universit\'es, CNRS UMR 7589,\\ 
        UPMC Paris 6, \\
        4 place Jussieu, 75005 Paris, France
\vskip0.29cm
      \end{itemize}}
\kern-3cm\vbox{
\begin{itemize}
 \item[$^2$] Albert Einstein Center for Fundamental Physics,\\
Institute for Theoretical Physics,\\ 
University of Bern,\\
Sidlerstrasse 5, 3012 Bern, Switzerland
      \end{itemize}\vskip0.05cm
}}
     \end{footnotesize}}
\end{center}
    \end{minipage}
    \vspace{0.5cm}\begin{minipage}{.7\linewidth}
\begin{center}     
{\it \begin{footnotesize}
\hbox{\kern4.cm\vbox{\vskip0cm
 \begin{itemize}
               \item[$^3$] Centre de Physique Th\'eorique,\\ 
        Ecole Polytechnique, CNRS UMR 7644,\\ 
        Universit\'e Paris-Saclay,\\
        91128 Palaiseau Cedex, France
\vskip0.29cm
      \end{itemize}}
}
     \end{footnotesize}}
\end{center}
     \end{minipage}
  \end{minipage}
}
\date{}

\title{\vspace{3.5cm}
 \boldmath \begin{Large}
    \textbf{Isometries, gaugings and ${\cal N}=2$ supergravity decoupling}
  \end{Large} \unboldmath
}

\begin{document}

\begin{titlepage}
\maketitle
\thispagestyle{empty}

 \vspace{-14.5cm}
  \begin{flushright}
  CPHT-RR040.072016
    \end{flushright}
 \vspace{12cm}

\begin{center}
\textsc{Abstract}\\  
\vspace{0.5cm}	
\begin{minipage}{1.0\linewidth}
We study off-shell rigid limits for the kinetic and scalar-potential terms of a single 
${\cal N}=2$ hypermultiplet. In the kinetic term, these rigid limits establish relations between four-dimensional 
quaternion-K\"ahler and hyper-K\"ahler target spaces with symmetry. The scalar potential is obtained by gauging
the graviphoton along an isometry of the quaternion-K\"ahler space. The rigid limits
unveil two distinct cases. A rigid ${\cal N}=2$ theory on Minkowski or on $\text{AdS}_4$ spacetime, depending on whether the 
isometry is translational or rotational respectively. We apply these results to the quaternion-K\"ahler 
space with $\text{Heisenberg}\ltimes U(1)$ isometry,
which describes the universal hypermultiplet at type-II string one-loop. 
\end{minipage}
\end{center}

\end{titlepage}

\onehalfspace

\vspace{-1cm}
\begingroup
\hypersetup{linkcolor=black}
\boldmath
\tableofcontents
\unboldmath
\endgroup
\noindent\rule{\textwidth}{0.6pt}

\section*{Introduction}
\addcontentsline{toc}{section}{Introduction}

The gravity decoupling is a subject of interest in supergravity and string theory. It is a precious source 
of information about the spectrum in models where supersymmetry is rigidly realized, controls the 
curvature of the background metric, and can shed light on the supersymmetry breaking mechanism. 

In the framework of local ${\cal N}=2$ supersymmetry, the hypermultiplet scalar dynamics is 
captured in $\sigma$-models with four-dimensional quaternion-K\"ahler target spaces \cite{BW}, and 
interaction potentials obtained by gauging some symmetries. Similarly, global ${\cal N}=2$ 
supersymmetry requires hyper-K\"ahler spaces \cite{AlvarezGaume:1981hm}. These are Ricci-flat
 K\"ahler spaces and for one hypermultiplet Riemann-self-dual. 

When supersymmetry is locally realized, the scalar curvature of the quaternion-K\"ahler space 
is directly proportional to the gravitational constant $k^2 =8\pi  M_{\text{Planck}}^{-2}$.
The decoupling limit consists in taking this coupling constant to zero \emph{i.e.} 
$M_{\text{Planck}}\to \infty$, which deforms the quaternion-K\"ahler geometry into a hyper-K\"ahler one. Such a 
limiting process must smoothly interpolate between the two 
geometries, and its description requires care. 
It implies to simultaneously ``zooming-in'' in order to recover 
non-trivial hyper-K\"ahler geometries, as performed in Ref.  \cite{AADT} for quaternion-K\"ahler spaces with
Heisenberg isometry (this symmetry was discussed in Ref. \cite{oneloop}). 

Bridging hyper-K\"ahler and quaternion-K\"ahler four-dimensional spaces has been discussed in some 
specific cases, including the 
quaternionic quotient method \cite{Galicki:1985qv,Galicki:1987jz}, or
the eight-dimensional hyper-K\"ahler cone technique \cite{deWit:1999fp,Anguelova:2004sj}.
Later, a general correspondence between
quaternionic manifolds with an isometry and 
hyper-K\"ahler manifolds with a rotational symmetry, endowed with a hyperholomorphic connection,
was found in \cite{Haydys} within a mathematical framework. This correspondence was further pursued  in \cite{Hitchin2012,Alekseevsky:2013nua}, and developed from a more physical perspective in \cite{Alexandrov:2011ac}. Finally progress has been made in the rigid limit of special quaternionic K\"ahler manifolds \cite{Gunara:2013rca},
constructed through the local c-map, reducing to hyper-K\"ahler spaces constructed by the rigid c-map \cite{Cecotti:1988qn,Ferrara:1989ik}.
It it obvious that several quaternion-K\"ahler spaces can have the same rigid limit. For instance, 
$SO(1,4)/SO(4)$ and $SU(1,2)/SU(2)\times U(1)$ lead to flat hyper-K\"ahler spaces. It is also true 
that one quaternion-K\"ahler space can have several rigid limits. A point that we discuss in this work.

More recently, a systematic pattern for connecting general quaternion-K\"ahler and hyper-K\"ahler 
spaces with symmetries was introduced in \cite{Antoniadis:2015egy}. The aim of the 
present article is to recast this method in more general terms, including in particular 
the scalar potential and its behaviour along the decoupling, its critical points and their 
supersymmetry properties, as well as the value of the cosmological constant. We will also 
discuss alternative decoupling limits, setting the control of the symmetry at the hyper-K\"ahler level.

We will first discuss hyper-K\"ahler and quaternion-K\"ahler spaces with symmetries, emphasizing a 
simple relationship between pairs of such spaces, which translates into the coupling or 
decoupling of gravity. This holds for the kinetic term of hypermultiplet scalars. The 
behaviour of the potential will be analyzed next, when this potential is obtained by 
gauging (for simplicity) the graviphoton along an isometry of the quaternion-K\"ahler space. 
Two separate regimes will be studied: the case where the decoupling of gravity leaves a 
rigid ${\cal N}=2$ theory on Minkowski background, and the alternative where the 
spacetime is $\text{AdS}_4$. For all these cases, we systematically study 
the mass spectrum.

We start with a short review on hyper-K\"ahler and quaternionic manifolds with a symmetry, Sec. \ref{HKQKiso}.
In Sec.  \ref{kineticrigid},  we investigate gravity decoupling limits of quaternionic manifolds with a symmetry, leading 
to hyper-K\"ahler spaces.
The analysis of the scalar potential by gauging the graviphoton is performed in Sec. \ref{gen.scalar.potential}. Finally, we study extensively in Sec. \ref{genquater} the quaternion-K\"ahler space with $\text{Heisenberg}\ltimes U(1)$ isometry and its decoupling limits. Two appendices follow, including the discussion of the
pseudo-Fubini--Study metric, which describes the universal hypermultiplet at string tree-level, Sec. \ref{PFSmetric}, 
and an alternative exhibition of generic, Ricci-flat, scalar-flat or Einstein four-dimensional K\"ahler spaces with a holomorphic isometry, Sec. \ref{Kahleriso}.

\boldmath
\section{Hyper-K\"ahler and quaternionic manifolds with a symmetry}\label{hyperQ}
\label{HKQKiso}
\unboldmath

A hyper-K\"ahler space in four dimensions is a K\"ahler manifold with self-dual Riemann tensor:
\begin{equation}
\label{self.duality}
R_{uvxy}-\frac12\,\varepsilon_{uv}^{\hphantom{uv}wz}\,R_{wz xy}=0\, .
\end{equation}
The indices $u,v,\ldots$ run from $1$ to $4$, and we have introduced  
$\varepsilon_{uvxy}=\sqrt{\det g}\, \epsilon_{uvxy}$ with $\epsilon_{0123}=1$. 
This space is Ricci-flat and endowed with 3 covariantly constant anti-self-dual 2-forms  
$J_K$. These form a triplet of $SU(2)$ complex structures normalized to satisfy
\begin{equation}
\label{identities}
\begin{split}
&\left(J_K\right)_u^{\hphantom{u}x}\,\left(J_L\right)_x^{\hphantom{x}v}=-\delta_{KL}\,\delta_u^v-\varepsilon_{KL}^{\hphantom{KL}M}\,\left(J_M\right)_u^{\hphantom{u}v}\,,\\
&\sum_{K=1}^3\,\left(J_K\right)_u^{\hphantom{u}v}\left(J_K\right)_x^{\hphantom{u}y}=g_{ux}\,g^{vy}-\delta_u^y\, \delta_x^v
+\varepsilon_{ux}^{\hphantom{ux}vy}\,.
\end{split}
\end{equation}

In the following, we will assume the existence of isometries.
As a consequence of Bianchi identity, a Killing vector $\xi$ satisfies 
\begin{equation}
\label{Killing.ident}
 \nabla_x\nabla_v\xi_u=R_{uvxy}\xi^y\,.
\end{equation}
Its 
(anti)-self-dual covariant derivatives\footnote{The  (anti)-self-dual components of a 2-form $A_{uv}$ are defined as $A_{uv}^{\pm}=\frac{1}{2}\left(A_{uv}\pm \frac{1}{2}\varepsilon_{uv}^{\hphantom{uv}wz} A_{wz}
\right)$.}
\begin{equation}
\label{self.cov}
k^\pm_{uv}=\frac12\left(\nabla_u \xi_v\pm
\frac{1}{2}\varepsilon_{uv}^{\hphantom{uv}wz}
\,\nabla_w \xi_z\right),
\end{equation}
obey  remarkable identities,
\begin{equation}
\label{propertieskpm}
\begin{split}
&g^{uv}\,k^\pm_{ux}\,k^\pm_{vy}=\frac{1}{4}\,g_{xy}\,k_\pm^2\,,\quad k_\pm^2:=k^\pm_{uv}k^{\pm uv}\,,\\
&g^{uv}\,k^\pm_{ux}\,k^\mp_{vy}=\frac12\left(\nabla_x\xi_z\nabla_y\xi^z-
\frac14\,g_{xy}\nabla_{u}\xi_{v}\nabla^{u}\xi^{v}\right),\\
&g^{uv}\,g^{xy}\,k^\pm_{ux}\,k^\mp_{vy}=0\,,
\end{split}
\end{equation}
valid irrespective of the nature of the space. 

The self-duality condition \eqref{self.duality} can be recast using \eqref{Killing.ident} as
\begin{equation}
\label{nablahk}
\nabla_x k^-_{uv}=0\,,
\end{equation} 
leading to
\begin{equation}
\partial_x\,k_-^2=0\quad\Longrightarrow\quad
k_-^2=c\,,
\end{equation}
where $c$ is a non-negative constant. 
Consequently, in hyper-K\"ahler spaces, a
Killing vector is translational 
if $k^-_{uv}=0$, and rotational otherwise 
\cite{Gibbons:1979zt, Boyer,GibbonsRubback}. 

In order to clarify the meaning of translational versus rotational isometry, we evaluate the Lie derivative on the complex structures with respect to the Killing vector $\xi$:
\begin{equation}
\label{LieHyper}
\left({\cal L}_\xi J_K\right)_{uv}=
\begin{cases}
\nabla_v\left(\left(J_K\right)_u^{\hphantom{u}w}\xi_w\right)-\nabla_u\left(\left(J_K\right)_v^{\hphantom{u}w}\xi_w\right)=
\partial_v\left(\left(J_K\right)_u^{\hphantom{u}w}\xi_w\right)-\partial_u\left(\left(J_K\right)_v^{\hphantom{u}w}\xi_w\right), & \\\nabla_u\xi_w \left(J_K\right)^w_{\hphantom{w}v}
 -\left(J_K\right)_u^{\hphantom{u}w}\nabla_w\xi_v=[\nabla\xi,J_K]_{uv}\,,
    \end{cases}
\end{equation}
where 
the bracket stands for the ordinary commutator of matrices (not to be confused with the Lie bracket).
The latter expression trivializes for  
a translational isometry \cite{GibbonsRubback}
\begin{equation}
\label{trans}
{\cal L}_\xi J_K=[k^+,J_K^-]=0\,,
\end{equation}
because $[A^-,B^+]=0$ for any pair of 2-forms $A$ and $B$.
Therefore a translational Killing vector $\xi$ is {\it triholomorphic}, leaving the 
three complex structures invariant. In addition, Eqs.~\eqref{LieHyper} and \eqref{trans} ensure the 
existence of a triplet of Killing potentials (moment maps) ${\cal K}_I$ defined as 
\begin{equation}
\left(J_I\right)_u^{\hphantom{u}w}\xi_w=-\frac12\,\partial_u{\cal K}_I\,.
\end{equation}

For a rotational isometry, we can always find a basis of complex structures such that
\begin{equation}
\label{rot}
{\cal L}_\xi J_1=J_2\,,\quad {\cal L}_\xi J_2=-J_1\,,\quad {\cal L}_\xi J_3=0\,.
\end{equation}
Consequently a rotational Killing vector $\xi$ is {\it simply holomorphic} since only one complex structure remains 
invariant. In addition, 
\eqref{LieHyper} and \eqref{rot} ensure the 
existence of a Killing potential ${\cal K}$ defined as 
\begin{equation}
\label{Killing.potential}
\left(J_3\right)_u^{\hphantom{u}w}\xi_w=-\frac12\,\partial_u{\cal K}\Longrightarrow
\xi_u=\frac{1}{2}\,\left(J_3\right)_u^{\hphantom{u}v}\,\partial_v{\cal K}\, ,
\end{equation}
and the hyper-K\"ahler metric $g^{\text{HK}}_{uv}$ satisfies the relation
\begin{equation}
\label{gHK.BK}
g^{\text{HK}}_{uv}=\frac{1}{2}\left(\delta_u^w\delta_v^z+
\left(J_\perp\right)_u^{\hphantom{v}w}\left(J_\perp\right)_v^{\hphantom{v}z}\right)\nabla_w\nabla_z{\cal K}\,,
\end{equation}
where $J_\perp$ is any complex structure orthogonal to $J_3$. Diagonalizing the latter selects
K\"ahler coordinates with
\begin{equation}
g^{\text{HK}}_{ab}=0\,,\quad g^{\text{HK}}_{a\bar b}=\nabla_a\nabla_{\overline b}{\cal K}=
\partial_a\partial_{\overline b}{\cal K}=
{\cal K}_{a\overline b}\,.
\end{equation}
Thus the Killing potential of $J_3$ is the K\"ahler potential for $J_\perp$ \cite{Hitchin:1986ea}.

Adapting a coordinate $\tau$ along the orbits of the Killing field as $\xi=\partial_\tau$, the hyper-K\"ahler metric reads:
\begin{equation}
\label{generalHK}
\begin{split}
&\text{d}s^2_{\text{HK}}=\frac1V\left(\text{d}\tau+\omega\right)^2+V\,\text{d}\ell^2\,,\\
&\text{d}\ell^2=\gamma_{ij}\,\text{d}X^i\text{d}X^j\,,\quad X^i=(X,Y,Z)\,.
\end{split}
\end{equation}
When $\xi$ is a translational Killing vector, we can use the Gibbons--Hawking frame \cite{Gibbons:1979zt} 
\begin{equation}
\label{GHmetric}
\gamma_{ij}=\delta_{ij}\,,\quad \nabla V=\nabla\times\omega\,.
\end{equation}
The complex structures read in this case:
\begin{equation}
\label{metric.complex}
J_1=-i\left(a\wedge b-\ov a\wedge\ov b\right),\quad
J_2=a\wedge b+\ov a\wedge\ov b\,,\quad
J_3=-i\left(a\wedge\ov a+b\wedge\ov b\right),
\end{equation}
with
\begin{equation}
\label{GHmetric.complex}
a=\frac{1}{\sqrt{2}}\left(\sqrt{V}\text{d}z+\frac{i}{\sqrt{V}}\left(\text{d}\tau+\omega\right)\right),\quad
b=\frac{1}{\sqrt{2}}\,\sqrt{V}\,\left(\text{d}x+i\,\text{d}y\right).
\end{equation}
Alternatively, for a rotational Killing vector, the Gibbons--Hawking frame \eqref{GHmetric} is traded 
for the Boyer--Finley one \cite{Boyer,GibbonsRubback}:\footnote{
Indices on scalar functions denote ordinary partial derivatives.} 
\begin{equation}
\begin{split}
\label{BFmetric}
&\text{d}\ell^2=\text{d}Z^2+\text{e}^\Psi\left(\text{d}X^2+\text{d}Y^2\right),\\
&V=\frac12\,\Psi_Z\,,\quad 
\omega=\frac12\,\Psi_Y\,\text{d}X-\frac12\,\Psi_X\,\text{d}Y\,,\\
&\Psi_{XX}+\Psi_{YY}+\left(\text{e}^\Psi\right)_{ZZ}=0\,.
\end{split}
\end{equation}
The complex structures are given by \eqref{metric.complex} with
\begin{equation}
\label{BFmetric.complex}
\begin{split}
a=\frac{1}{\sqrt{2}}\left(\sqrt{V}\text{d}z+\frac{i}{\sqrt{V}}\left(\text{d}\tau+\omega\right)\right),\quad
b=\frac{\text{e}^{i\,\tau}}{\sqrt{2}}\,\sqrt{V}\,\text{e}^{\nicefrac{\Psi}{2}}\left(\text{d}x+i\,\text{d}y\right).
\end{split}
\end{equation}
Notice the explicit $\tau$-dependence, necessary for \eqref{rot} to hold. Finally, the Killing potential $\mathcal{K}$ for 
$\xi=\partial_\tau$ is obtained using
\eqref{Killing.potential}, \eqref{metric.complex} and \eqref{BFmetric.complex}:
 \begin{equation}
 \label{Kcal.gen}
 {\cal K}=2Z\,.
 \end{equation}

We now turn to quaternion-K\"ahler spaces. These are Einstein 
and conformally self-dual:\footnote{\label{Itoh} 
Four-dimensional quaternion-K\"ahler spaces are in general not K\"ahler, but there are exceptions.
These include $SU(3)/SU(2)\times U(1)$ and $SU(1,2)/SU(2)\times U(1)$ (see  App.~\ref{PFSmetric}).
The K\"ahler structure introduces a canonical orientation and
self-duality is not equivalent with anti-self-duality. In particular, 
a four-dimensional K\"ahler metric which is Weyl {\it anti}-self-dual has
vanishing scalar curvature \cite{Itoh}.}

\begin{equation}
\label{WSD}
W_{uvxy}-\frac12\,\varepsilon_{uv}^{\hphantom{uv}wz}\,W_{wz xy}=0\,.
\end{equation}
We will here normalize the scalar curvature to $R=-12$.  
Assuming again the existence of an isometry and using Eq.~\eqref{Killing.ident}, the Einstein-space condition $R_{uv}=-3g_{uv}$ and \eqref{WSD}, we find:
\begin{equation}
\label{nablakm}
\nabla_x k^-_{vu}=-
\frac12\,\left(g_{ux}g_{vy}-g_{uy}g_{vx}-\varepsilon_{uvxy}\right)\xi^y\,,
\end{equation}
instead of \eqref{nablahk}.
Contrary to what happens for hyper-K\"ahler spaces, the distinction between translational and 
rotational Killings is no longer relevant here, and the Gibbons--Hawking or the Boyer--Finley forms 
\eqref{generalHK}, \eqref{GHmetric} and \eqref{BFmetric} are replaced by the Przanowski--Tod 
frame \cite{prz0,prz1,Tod,Tod:2006wj}, where\footnote{The integrability condition 
for $\omega$, given by the linearized Toda equation
$
\left(\partial_X^2+\partial_Y^2\right)V+\partial_Z^2\left(V\text{e}^\Psi\right)=0
$, is actually a consequence of the last equations in \eqref{prztod1}.}
\begin{equation}
\begin{split}
\label{prztod1}
&\text{d}s_{\text{QK}}^2=\frac{1}{Z^2}\left(\frac{1}{V}(\text{d}\tau+\omega)^2+
V\left(\text{d}Z^2+\text{e}^\Psi\left(\text{d}X^2+\text{d}Y^2\right)\right)\right),\\
&\text{d}\omega= V_X \, \text{d}Y\wedge \text{d}Z+ V_Y \, \text{d}Z\wedge \text{d}X+
\left(V\, \text{e}^\Psi\right)_Z\,\text{d}X\wedge \text{d}Y\,,\\
&\Psi_{XX}+\Psi_{YY}+\left(\text{e}^\Psi\right)_{ZZ}=0\,,\quad
2 V=2-Z\, \Psi_Z\,.
\end{split}
\end{equation}
A straightforward computation shows that the coordinate $Z$ is  related to 
the anti-self-dual covariant derivative of the Killing field $\xi=\partial_\tau$:
\begin{equation}
\label{kmZ0}
\frac{1}{Z^2}
=k_-^2=k^-_{uv}k^{-uv}\,.
\end{equation}

\boldmath
\section{The kinetic term and the rigid limits}
\label{kineticrigid}
\unboldmath

As discussed in Ref.~\cite{Antoniadis:2015egy}, using any solution of the Toda equation,
one can build both a quaternion-K\"ahler space with symmetry, expressed \emph{\`a la} Przanowski--Tod, 
and  a hyper-K\"ahler space in the Boyer--Finley frame. This sets a simple one-to-one relationship among these spaces. 

Although this relationship sounds formal, it supports a deeper geometric interpretation:
the hyper-K\"ahler member of the pair appears actually as a zooming-in of the quaternionic member around the fixed point 
of the isometry that supports the fiber in the Przanowski--Tod frame. This isometry survives in the hyper-K\"ahler space 
as a simply holomorphic symmetry. From a physical viewpoint, as we will see soon, the limiting procedure at hand goes along with the gravity decoupling limit.

In order to elaborate on the geometric picture of the above correspondence, 
we recall that the kinetic term reads:\footnote{
Out of the full supergravity action, we consider the following part
\begin{equation*}
S=\int\sqrt{-g}\,\text{d}^4x\,\left(\mathscr{K}-\mathscr{V}+{\mathscr L}_{\text{EH}}\right)\,,\quad
{\mathscr L}_{\text{EH}}=\frac{1}{k^2}\left(\frac{{\cal R}}{2}-\Lambda\right)\,,
\end{equation*}
where $\mathscr{K}$, $\mathscr{V}$ and $\mathscr{L}_\text{EH}$ correspond to the kinetic, potential and Einstein--Hilbert 
terms respectively.
}
\begin{equation}
\label{kinetic}
\mathscr{K}=\frac{1}{2k^2}\,G^{\mu\nu} g^{\text{QK}}_{uw}\,\partial_\mu q^u\,\partial_\nu q^w\,.
\end{equation}
Here, $G_{\mu\nu}$ is the spacetime metric and $x^\mu$ the associated coordinates, whereas $g^{\text{QK}}_{uv}$ 
is the quaternionic target-space, coordinated with $q^u$.
Notice that  $q^u, g^{\text{QK}}_{uv}$ and $G_{\mu\nu}$ are dimensionless, whereas $\partial_\mu$ 
have dimension one and $\mathscr{K}$ dimension four.

The gravity decoupling limit of the quaternion-K\"ahler space, reached at $k\to0$, should be taken in a 
zoom-in manner in order to avoid the trivialization of the geometry into flat space. In that aim 
we introduce the following redefinitions:
\begin{equation}
\begin{split}
\label{zoomtrans}
Z=\alpha\widehat Z-\delta\,,\quad V= \delta\, \widehat V\,,\quad &\tau= \alpha\,\delta\, \widehat\tau\,,\quad 
\omega= \alpha\,\delta\,\widehat \omega\,,\\
X=\alpha \widehat X\,,\quad Y=&\alpha \widehat Y\,,\quad \Psi=\alpha\widehat\Psi\,,
\end{split}
\end{equation}
in the Przanowski--Tod metric \eqref{prztod1}, leading to
\begin{equation}
\label{prztod2}
\begin{split}
&\text{d}s_{\text{QK}}^2=\frac{\alpha^2\,\delta}{(\alpha\,\widehat Z-\delta)^2}\left(\frac{1}{\widehat V}
\left(\text{d}\widehat\tau+\omega\right)^2+\widehat V\left(\text{d}\widehat Z^2+\text{e}^{\alpha\widehat\Psi}
\left(\text{d}\widehat X^2+\text{d}\widehat Y^2\right)\right)\right),\\
&\text{d}\omega= V_{\widehat X} \, \text{d}\widehat Y\wedge \text{d}\widehat Z+
 V_{\widehat Y} \, \text{d}\widehat Z\wedge \text{d}\widehat X
+\left(V\, \text{e}^{\alpha\widehat\Psi}\right)_{\widehat Z}\text{d}\widehat X\wedge \text{d}\widehat Y\, , \\
&\widehat V=\frac{1}{2}\,\Psi_{\widehat Z}+\frac{1}{2\delta}\left(2-\alpha\widehat Z\,\Psi_{\widehat Z}\right),
\end{split}
\end{equation}
and 
\begin{equation}
\label{kmZ}
k_-^2=k^-_{uv}k^{-uv}=\frac{\alpha^2\delta^2}{(\alpha\widehat Z-\delta)^2}\,.
\end{equation}

\paragraph{From Przanowski--Tod to Boyer--Finley}
The kinetic term \eqref{kinetic} with the insertion of \eqref{prztod2},
remains finite in the double-scaling limit
\begin{equation}
\label{zoomin}
\alpha=1\,,\quad k \to0\,,\quad \delta\to\infty\,,\quad k^2\,\delta=\frac{1}{\tilde\mu^2}\,,
\end{equation}
where $\tilde\mu$ is an arbitrary finite mass scale. In this limit, \eqref{kinetic} reads:
\begin{equation}
\label{kineticHK}
\mathscr{K}=\frac{\tilde\mu^2}{2}\,G^{\mu\nu} g^{\text{HK}}_{uw}\,\partial_\mu q^u\,\partial_\nu q^w\,,
\end{equation}
where $g^{\text{HK}}_{uv}$ is the hyper-K\"ahler space in Boyer--Finley frame \eqref{generalHK} and 
\eqref{BFmetric}, corresponding to the solution of the Toda equation used in the quaternionic metric $g^{\text{QK}}_{uw}$ of 
 \eqref{kinetic}.  
Furthermore, using \eqref{kmZ} we find that $k_-^2$ remains non-vanishing
in the double-scaling limit \eqref{zoomin}.
Thus,  the original quaternionic isometry is mapped onto a simply holomorphic Killing vector $\partial_\tau$.
Other isometries of the quaternionic metric also survive in the hyper-K\"ahler limit, if they 
commute with $\partial_\tau$. Additional isometries may also exist.

Several remarks are in order here regarding the implementation of the double-scaling limit \eqref{zoomin}.
This limit consists of zooming-in around $Z\to -\infty$ in the quaternion-K\"ahler space. 
The latter is described, in the Przanowski--Tod representation, by a  solution $\Psi(X,Y,Z)$ of 
the Toda equation, and $Z\to -\infty$ is the fixed locus of the isometry generated by $\partial_\tau$, 
as the norm of this Killing vanishes in that limit.  
The double-scaling limit \eqref{zoomin} is consistent provided 
${\widehat\Psi}(\widehat X,\widehat Y,\widehat Z)
= \frac{1}{\alpha} \Psi(\alpha \widehat X,\alpha\widehat Y,\alpha\widehat Z-\delta)$ introduced in \eqref{zoomtrans}
makes sense when $\alpha\to 1$ and $\delta \to \infty$. Being a solution of a partial 
differential equation, $\Psi(X,Y,Z)$ is actually a function of $X+X_0\,, Y+Y_0\,, Z+Z_0$, where $X_0\,,Y_0\,, Z_0 $ 
are arbitrary constants. This freedom makes it possible to tune $Z_0$ so as to absorb $\delta$ before the 
limit is taken. In this way, the limit does not affect  $\Psi(X,Y,Z)$ and the very same function can thus 
be used on the two sides of the limit. This is why the double-scaling limit under consideration is equivalent 
to the one-to-one correspondence among quaternionic and hyper-K\"ahler spaces with symmetry set in 
the beginning of the present 
section, and based on the use of a given solution of the Toda equation. From this perspective, the relationship at hand
can either be interpreted as a decoupling of gravity when starting from a quaternion-K\"ahler space, or as a coupling 
to gravity, when starting from a hyper-K\"ahler $\sigma$-model endowed with a simply holomorphic 
Killing vector  sustaining a rigid ${\cal N}=2$ model.

It should finally be stressed that the double-scaling limit under investigation can be taken 
for any isometry of the quaternion-K\"ahler space. This provides as many decoupling limits of gravity 
as symmetries in the $\sigma$-model, not all inequivalent though.  
We will come back to that in the examples of Sec. \ref{genquater} and App. \ref{PFSmetric}, 
when discussing in particular the fate of the symmetries along the decoupling.

\paragraph{From Przanowski--Tod to Gibbons--Hawking}

Starting with a quaternion-K\"ahler space with symmetry in the Przanowski--Tod representation, 
we can reach in some cases another hyper-K\"ahler space. This zoom-in limit is not necessarily taken in the
same area of the manifold as the previous limit.\footnote{See the discussion at the end of Section 
\ref{seclimits}.}
Instead of the double-scaling limit \eqref{zoomin}, 
we perform the following triple-scaling limit on
the kinetic term \eqref{kinetic} with redefinitions \eqref{zoomtrans}:
\begin{equation}
\label{zoomin1}
k \to0\,,\quad \alpha\to0\,,\quad \delta\to\infty\,,\quad \frac{k^2\,\delta}{\alpha^2}=\frac{1}{\tilde\mu^2}\,.
\end{equation}
The kinetic term is still given by \eqref{kineticHK}, where now 
 $g^{\text{HK}}_{uv}$ is a hyper-K\"ahler space in the Gibbons--Hawking frame \eqref{generalHK}, 
\eqref{GHmetric}. Hence, the original quaternionic isometry generated by $\partial_\tau$ becomes 
triholomorphic in the hyper-K\"ahler limit, where indeed $k_-^2$ and $k^-_{uv}$ vanish (see \eqref{kmZ}).
Again, additional isometries may exist, as in the previous double-scaling limit. 
 
The rigid limit under consideration is again a zooming-in around  $Z \to -\infty$, where the norm of 
$\partial_\tau $ vanishes, further restricted to $X=Y=0$. 
The limiting hyper-K\"ahler space exists as long as
$
\widehat V=\frac{1}{2\alpha}\,\partial_{\widehat Z}\Psi(\alpha\widehat X, 
\alpha\widehat Y,\alpha\widehat Z)
$
makes sense when $\alpha \to 0$.  This requirement sets restrictions to the Przanowski--Tod 
geometries that allow for the triple-scaling limit, contrary to the previous double-scaling 
limit, which always exists. Despite that, non-trivial examples can be successfully worked out. 
In conclusion, although the 
original Przanowski--Tod isometry is generically simply holomorphic in the limiting 
hyper-K\"ahler space, the option for a triholomorphic limit exists in certain instances. 
 
 \paragraph{Alternative limit to Gibbons--Hawking}

Before closing this section, we would like to mention an alternative possibility for reaching 
a hyper-K\"ahler Gibbons--Hawking space from a Przanowski--Tod quaternionic geometry. 
This limit is peculiar, because when it exists, it always leads to the same space, namely 
the unique hyper-K\"ahler invariant under 
$\text{Heisenberg}\ltimes U(1)$ symmetry \cite{BEPS, Antoniadis:2015egy}.

Assume that a line $(X_0, Y_0, Z_0)$ exists in a quaternion-K\"ahler space of the Przanowski--Tod 
type \eqref{prztod1}, such that 
\begin{equation}
\label{alter.condit}
\begin{cases}
V(X_0, Y_0, Z_0)=0\Leftrightarrow \left.Z\partial_Z\Psi\right\vert_0=2\, ,\\
\omega_X\vert_0=\omega_Y\vert_0=\omega_Z\vert_0=0\, ,\\
\Psi(X_0, Y_0, Z_0)=\psi_0\, ,\quad \partial_Z\left(Z\partial_Z\Psi\right)\vert_0=-2\psi_Z\, ,\\
\partial_X\left(Z\partial_Z\Psi\right)\vert_0=-2\psi_X\,  \text{e}^{\nicefrac{\psi_0}{2}}\, ,\quad
\partial_Y\left(Z\partial_Z\Psi\right)\vert_0=-2\psi_Y\,  \text{e}^{\nicefrac{\psi_0}{2}}\, ,
\end{cases}
\end{equation}
with $\psi_0, \psi_X, \psi_Y$ and $ \psi_Z$ finite, potentially vanishing constants.
In the neighborhood of this line, we define the coordinates $\widehat{X} ,\widehat{Y} ,\widehat{Z}$:
\begin{equation}
\label{alter.condit1}
X= X_0+\left(k\tilde\mu\right)^{\nicefrac{2}{3}}\,\text{e}^{-\nicefrac{\psi_0}{2}}\, \widehat X\,,\quad
Y= Y_0+\left(k\tilde\mu\right)^{\nicefrac{2}{3}}\,\text{e}^{-\nicefrac{\psi_0}{2}}\,\widehat Y\,,\quad
Z= Z_0+\left(k\tilde\mu\right)^{\nicefrac{2}{3}}\,\widehat Z\,,
\end{equation}
and we expand $V$ and $\omega$ at linear order, since we are ultimately interested in the 
scaling limit $k\to 0$. We find that
 \begin{equation}
 \begin{split}
&V\approx\left(k\tilde\mu\right)^{\nicefrac{2}{3}}\left(
\psi_X\,  \widehat X + \psi_Y\,  \widehat Y +\psi_Z\,  \widehat Z 
\right)=\left(k\tilde\mu\right)^{\nicefrac{2}{3}} \widehat V\,,\\
&\omega\approx\left(k\tilde\mu\right)^{\nicefrac{4}{3}}\left(
\psi_X\,  \widehat Y\, \text{d}\widehat Z + \psi_Y\,  
\widehat Z\, \text{d}\widehat X +\psi_Z\,  \widehat X \, \text{d}\widehat Y
\right)=\left(k\tilde\mu\right)^{\nicefrac{4}{3}} \widehat \omega\,.
 \end{split}
\end{equation}
Introducing finally 
\begin{equation}
\label{alter.condit2}
\tau=\left(k\tilde\mu\right)^{\nicefrac{4}{3}}\,\widehat\tau\, ,
\end{equation}
we can proceed with the decoupling limit $k\to 0$ in the kinetic term 
\eqref{kinetic} and we find the rigid limit  \eqref{kineticHK} with the 
hyper-K\"ahler metric  
\begin{equation}
\label{qydjfosls}
\text{d}s_\text{HK}^2=\frac{1}{Z_0^2}\left(\frac{1}{\widehat V}\left(\text{d}\widehat\tau+\widehat\omega\right)^2
+\widehat V\left(\text{d}\widehat X^2+\text{d}\widehat Y^2+\text{d}\widehat Z^2\right)\right),
\end{equation}
where both $\widehat V$ and $\widehat \omega$ are linear in the hated coordinates with
\begin{equation}
\label{GBcons}
\widehat\nabla\, \widehat V=\widehat\nabla\times\widehat\omega=\{\psi_X,\psi_Y,\psi_Z\}\, .
\end{equation}

Thanks to the relation \eqref{GBcons}, it is always possible to trade the coordinates 
$\{\widehat \tau, \widehat X,\widehat Y, \widehat Z\}$ for $\{t,x,y,z\}$:
\begin{equation*}
\begin{split}
&t=Z_0^{\nicefrac{-2}{3}}\,r_0^{\nicefrac{1}{3}}\,\left(\cos\vartheta_0\,\widehat Z+\sin\vartheta_0
\left(\cos\varphi_0\,\widehat X+\sin\varphi_0\,\widehat Y\right)\right),\\
&x=Z_0^{\nicefrac{-2}{3}}\,r_0^{\nicefrac{1}{3}}\,\left(-\sin\vartheta_0\,\widehat Z+\cos\vartheta_0
\left(\cos\varphi_0\,\widehat X+\sin\varphi_0\,\widehat Y\right)\right),\\
&y=Z_0^{\nicefrac{-2}{3}}\,r_0^{\nicefrac{1}{3}}\,\left(-\sin\varphi_0\,\widehat X+\cos\varphi_0\,\widehat Y\right),\\
&z=Z_0^{\nicefrac{-4}{3}}\,r_0^{\nicefrac{-1}{3}}\,\left(\widehat\tau+f\right),
\end{split}
\end{equation*}
where
\begin{equation*}
f=\frac14\,r_0\,\cos\vartheta_0\,\sin2\varphi_0\,\left(\widehat X^2-\widehat Y^2\right)+
r_0\,\widehat Y\left(\sin^2\varphi_0\cos\vartheta_0\widehat X+\cos\varphi_0\sin\vartheta_0\widehat Z\right)
\end{equation*}
and $(r_0,\vartheta_0,\varphi_0)$ are constants defined as
\begin{equation*}
\psi_X=r_0\sin\vartheta_0\cos\varphi_0\,,\quad \psi_Y=r_0\sin\vartheta_0\sin\varphi_0\,,\quad \psi_Z=r_0\cos\vartheta_0.
\end{equation*}
The metric thus reads: 
\begin{equation}
\label{jdkdkfjh}
\text{d}s_{\text HK}^2 =\frac{1}{t} (\text{\text{d}}z + x \, \text{\text{d}}y)^2 + 
t\left(\text{\text{d}}t^2 + \text{\text{d}}x^2 +\text{\text{d}}y^2\right),
\end{equation}
which is the unique hyper-K\"ahler space invariant under 
$\text{Heisenberg}\ltimes U(1)$ symmetry \cite{BEPS, Antoniadis:2015egy}, 
generated by $(\mathscr{X}, \mathscr{Y}, \mathscr{Z}, \mathscr{M})$ obeying
\begin{equation}
\label{Heisenbhyper}
\left[\mathscr{X}, \mathscr{Y}\right]=\mathscr{Z},\quad
\left[\mathscr{M}, \mathscr{X}\right]=\mathscr{Y}\,,
\quad\left[\mathscr{M}, \mathscr{Y}\right]=-\mathscr{X}\,,
\end{equation}
and realized as
\begin{equation}
\label{Heisenbhyper1}
\begin{split}
&\mathscr{X}=\partial_x-y \partial_z\,, \quad
\mathscr{Y}=\partial_y\,,\quad
\mathscr{Z}=\partial_z\,,\quad 
\mathscr{M}=y\, \partial_x-x\, \partial_y+\frac{1}{2}\left(x^2-y^2\right) \partial_z\, .
\end{split}
\end{equation}
The Killing fields $\mathscr{X}, \mathscr{Y}, \mathscr{Z}$ are  triholomorphic (translational) whereas $\mathscr{M}$ is  simply holomorphic (rotational).

If conditions \eqref{alter.condit} are met, the gravity decoupling limit under 
consideration provides the specific hyper-K\"ahler space \eqref{jdkdkfjh}. This occurs for 
example in the two-parameter family of $U(1)\times U(1)$-symmetric quaternion-K\"ahler spaces obtained 
by quaternionic quotient based upon gauging $\mathscr{Y}$ and $\mathscr{Z}$
inside the $Sp(2, 4)$ of the ${\cal N} = 2$ hypermultiplet manifold \cite{Antoniadis:2015egy}. 
This family of quaternion-K\"ahler spaces contains the sub-family of the Heisenberg $\ltimes U(1)$ spaces
resulting from a $\mathscr{Z}$ gauging \cite{AADT}.

Finally, it is useful to exhibit K\"ahler coordinates for the hyper-K\"ahler space \eqref{jdkdkfjh}.
There are at least two inequivalent K\"ahler  coordinate systems adapted to the isometry at hand.
In the first one, the action of $\mathscr{M}$ is not holomorphic:
\begin{equation}
\label{KahlerM1}
\begin{split}
&\Phi=t+i\, y\,,\quad T=-t\, x+i\, z\,,\quad 
K=\frac{\left(T+\ov T\right)^2}{\Phi+\ov\Phi}+\frac{1}{12}\,\left(\Phi+\ov\Phi\right)^3\,,\\
&\mathscr{X}=-\Phi\,\partial_T-\ov\Phi\,\partial_{\ov T}\,,\quad 
\mathscr{Y}=i\,\left(\partial_\Phi-\partial_{\ov\Phi}\right),\quad
\mathscr{Z}=i\,(\partial_T-\partial_{\ov T})\,,\\
&\mathscr{M}=\frac{1}{2i}\left(\Phi-\ov\Phi\right)\mathscr{X}
+\frac{T+\ov T}{\Phi+\ov\Phi}\,\mathscr{Y}+
\frac12\,\left(\left(\frac{T+\ov T}{\Phi+\ov\Phi}\right)^2-\frac14\,\left(\Phi-\ov\Phi\right)^2\right)\mathscr{Z}\,,
\end{split}
\end{equation}
while it is holomorphic in the second:
\begin{equation}
\begin{split}
&\Psi=x+i y\,,\quad U=\frac14\left(2 t^2-x^2-y^2\right) +i \left(\frac12 x\,y + z\right),\\
&Q=U+\ov U+\frac12\,\Psi\ov\Psi=t^2\,,\quad K=\frac43\,Q^{\nicefrac{3}{2}}\,,\\
&\mathscr{X}=-\frac12\,\Psi\partial_U-\frac12\,\ov\Psi\partial_{\ov U}+\partial_\Psi+\partial_{\ov\Psi}\,,
\quad \mathscr{Y}=\frac{i}{2}\,\left(\Psi\partial_U-\ov\Psi\partial_{\ov U}\right)+i\,(\partial_\Psi-\partial_{\ov\Psi}),\\
&\mathscr{Z}=i\left(\partial_U-\partial_{\ov U}\right),\quad
\mathscr{M}=-i\left(\Psi\partial_\Psi-\ov\Psi\partial_{\ov\Psi}\right).
\end{split}
\end{equation}

\boldmath
\section{The scalar potential}
\label{gen.scalar.potential}
\unboldmath
\subsection{Potential and spectrum}
\label{gen.scalar.potential.cs}
The scalar potential of ${\cal N}=2$ supergravity theories is obtained by gauging one or several symmetries realized as isometries on the quaternion-K\"ahler geometry. These isometries act on the components of hyper- and vector multiplets. The gauging procedure involves in general the graviphoton and possibly other gauge fields in vector multiplets. Here, we will gauge isometries of the hypermultiplet $\sigma$-model using only the graviphoton as gauge field. Despite the obvious limitations of such a choice (e.g. partial breaking into ${\cal N}=1$ is impossible
without vector multiplets: the massive ${\cal N}=1$ gravitino multiplet includes two massive spin-one fields), its analysis is rich and instructive, as we will see in the following. When considering extra vector multiplets, the output of the gauging depends on whether the isometry acts or not on the vectors at hand: when it does not, one commonly obtains a run-away behaviour, alternatively the scalar potential is more intricate and no generic conclusion can be drawn \emph{a priori}. We leave this investigation for the future.

In ${\cal N}=2$ supergravity, the choice of the symmetry to be gauged is free. In particular, the concept of translational versus rotational isometries is not pertinent in the quaternion-K\"ahler target space. This is in contrast with the global-supersymmetry case, and one of our purposes is to analyze how this distinction emerges in the gravity-decoupling limit. As we will see, it is intimately linked to the background spacetime geometry 
(Minkowski or AdS) dictated by the scalar potential.

The gauging procedure works as follows. The hypermultiplet metric is quaternion-K\"ahler and admits
three complex structures satisfying the quaternionic algebra \eqref{identities}. For each isometry generated by a Killing field $\xi$, one defines the corresponding Killing prepotentials  \cite{D'Auria:2001kv}:
\begin{equation}
\label{prepotential.quat}
P_I=-\frac{1}{4k^2}\, (J_I)^{u}_{\hphantom{u}v}\,\nabla_u\xi^v\,. 
\end{equation}
Once an isometry is selected, its gauging with the graviphoton produces a superpotential $W$ expressed in terms of the Killing prepotentials as:
\begin{equation}
\label{superPT}
W^2=\sum_{I=1}^3\,P_I P_I\,.
\end{equation}
The corresponding scalar potential $\mathscr{V}_\xi$ takes the form \cite{Ceresole:2001wi}
\begin{equation}
\label{potentialgraviI}
\mathscr{V}_\xi=k^2|X^0|^2\,\left(-6W^2+4g^{uv}\partial_u W\partial_v W\right).
\end{equation}
In the latter expression, $X^0$ is the compensator for dilation symmetry, gauge-fixed to $\nicefrac{1}{k}$.

Using \eqref{prepotential.quat}, \eqref{superPT}, the quaternion algebra \eqref{identities} 
and some of the identities \eqref{propertieskpm}, we find (the gravitational coupling $k^2$ should not be confused with the square of the anti-self-dual covariant derivative of the Killing $k_-^{2}$)
\begin{equation}
\label{W2}
W^2=\frac{1}{4k^4}\,k_-^{2}\,,
\end{equation}
and its derivative\footnote{We have used \eqref{nablakm}: $\partial_x k_-^{2}=
\nabla_x \left(k^-_{uv}k^{-uv}\right)=4k^-_{xy}\,\xi^y
$.}
\begin{equation}
\label{derivW}
\partial_x W^2=\frac{1}{k^4} k^-_{xy}\,\xi^y\,.
\end{equation}
Hence, we retrieve
\begin{equation}
\label{partialW}
g^{uv}\partial_u W\partial_v W=\frac{1}{4W^2}\,g^{uv}\partial_u W^2\partial_v W^2=
\frac{1}{k^4k_-^2}\,g^{uv}\,k^-_{ux}\,k^-_{vy}\xi^x\xi^y=\frac{1}{4k^4}\,g_{uv}\,\xi^u\xi^v\,.
\end{equation}
Expressions  \eqref{W2} and \eqref{partialW} enable us to produce the following equivalent expressions for the scalar potential \eqref{potentialgraviI}:  
\begin{equation}
\boxed{
\label{potentialgravi}
\mathscr{V}_\xi=\frac{1}{k^4}\begin{cases}
k^4\left(-6 W^2+ 4\,g^{uv}\,\partial_u W\partial_v W\right), & \\
 -6\,k^4 \sum\limits_{I=1}^3\,P_I P_I+ g_{uv}\,\xi^u\xi^v\,, & \\
      -\frac{3}{2}\,k_-^2+ g_{uv}\,\xi^u\xi^v\,.
    \end{cases}
}
\end{equation}

Given the scalar potential \eqref{potentialgravi}, it is straightforward to investigate its supersymmetric vacuums.  Those obey $\langle\partial_u W\rangle=0\Leftrightarrow\langle\xi^u \rangle=0$ (or equivalently $\langle g_{uv}\xi^u\xi^v \rangle=0$ -- the brackets mean as usual that the quantity under consideration is evaluated at the vacuum). 
The cosmological constant
is related to the -- generically non-vanishing -- value of the potential at the extremum, 
\begin{equation}
\label{cosmological}
\frac{1}{k^2}\,\Lambda=\langle \mathscr{V}_\xi\rangle=-\frac{3}{2k^4}\,\langle k_-^2\rangle.
\end{equation}
Expanding around the vacuum allows to determine the mass matrix:
\begin{equation}
\label{Mass}
M^u_{\hphantom{u}v}=\frac{k^2}{2}\,\langle\, g^{ux}\,\partial_x\partial_v\,\mathscr{V}_\xi\rangle=
\frac{1}{k^2}\left\langle
k^{+ uw}
k^+_{vw}
- 2
k^{- uw}
k^-_{vw}
-
k^{+ uw}
k^-_{vw}
\right\rangle .
\end{equation}
Equation \eqref{cosmological} shows that the spacetime has negative curvature ${\cal R}=4\Lambda$, whereas Eq.~\eqref{Mass} describes the spectrum of an ${\cal N}=2$ chiral multiplet 
in $\text{AdS}_4$ 
\cite{Breitenlohner:1982bm,Breitenlohner:1982jf}. Indeed, using the identities \eqref{propertieskpm} 
the mass matrix can be diagonalized with eigenvalues
\begin{equation}
\label{genmasspec}
\boxed{
\begin{split}
M_A^2=m^2-2\mu^2-m\mu\,,&\quad M_B^2=m^2-2\mu^2+m\mu\,,\\
m^2= \frac{1}{2k^2}\,\langle k_+^2\rangle\,,\quad &\mu^2= \frac{1}{2k^2}\,\langle k_-^2\rangle=-\frac{\Lambda}{3}\,.
\end{split}
}
\end{equation}
Recapitulating, $\langle k_+^2\rangle$ and $\langle k_-^2\rangle$
control respectively the physical mass $m$ of the 
chiral multiplet and the cosmological constant $\Lambda$ of the anti-de Sitter spacetime. The described vacuum is stable 
as it is supersymmetric. Furthermore it satisfies the 
Breitenlohner--Freedman stability bound \cite{Breitenlohner:1982bm,Breitenlohner:1982jf}
\begin{equation}
M_{A,B}^2\geqslant\frac34\,\Lambda\quad\Longrightarrow\quad \left(m\pm\frac{\mu}{2}\right)^2\geqslant0\, .
\end{equation}
In \eqref{genmasspec}, $M^2_{A,B}$ is the coefficient of the Lagrangian mass term. A field with a shift symmetry
or a flat direction of the potential correspond to $M^2_A=0$ or $M^2_B=0$. This does not mean that the field
is massless. In $\text{AdS}_4$ space, a field propagating on the lightcone has a Lagrangian mass term 
with coefficient $\nicefrac{2\Lambda}{3}=-2\mu^2$. We may then have a hypermultiplet with two flat directions and two
massless scalars, if $m=\pm\mu$.

\subsection{The decoupling limits}
\label{gen.scalar.potential.dc}
So far our analysis has been confined in the supergravity framework, \emph{i.e.} with coupling to gravity. We would like now to investigate the decoupling limit, and in particular the behaviour of the above scalar potential \eqref{potentialgravi} in the rigid limits presented in Sec. \ref{kineticrigid}. In these decoupling limits as they emerge in the analysis of the kinetic term, the Killing field supporting the fiber of the quaternion-K\"ahler space in its Przanowski--Tod representation \eqref{prztod1},  $\partial_\tau$, plays a preferred role. In particular, the zooming-in is triggered around distinguished points, where the norm of this vector is singular. However, in general, this specific Killing field needs not be the one that enters the gauging procedure. Consequently, the behaviour of the potential in the decoupling limits associated with the kinetic term is not unique and also depends on the isometry chosen for gauging. 

In the present general analysis, we do not make any assumption regarding extra isometries in the quaternion-K\"ahler space. Hence, we limit our discussion to the gauging of the graviphoton along the shift isometry $\xi=g\,\partial_\tau$ fibering the Przanowski--Tod metric \eqref{prztod1}, with $g$ a dimensionless gauge coupling.  
In Sec. \eqref{genquater}, when dealing with a specific quaternionic space, we will use the option of gauging isometries other than the one carrying the fiber and being responsible for the decoupling in the kinetic term.

Using \eqref{kmZ0} with $\xi=g\,\partial_\tau$, we find for the scalar potential \eqref{potentialgravi}: 
\begin{equation}
\label{potentialgraviPT}
\mathscr{V}_\xi=\frac{g^2}{k^4}\,\left(-\frac{3}{2Z^2}+\frac{1}{Z^2V}\right).
\end{equation}
Our goal is to investigate the behaviour of the latter expression when $k\to0$. As already discussed extensively in Sec. 
\ref{kineticrigid} for the kinetic term, this rigid limit must be taken in a zoom-in manner. Performing the redefinitions as stated in \eqref{zoomtrans}, we obtain:
\begin{equation}
\label{potentialgravitransf}
\mathscr{V}_\xi=\frac{\widehat g^2\alpha^2\,\delta^2}{k^4}\,
\left(-\frac{3}{2(\alpha\,\widehat Z-\delta)^2}+\frac{1}{(\alpha\,\widehat Z-\delta)^2\delta \widehat V}\right),\quad
\widehat g=\frac{g}{\alpha\,\delta}\,.
\end{equation}

\paragraph{From Przanowski--Tod to Boyer--Finley}
Performing the double-scaling limit \eqref{zoomin} on \eqref{potentialgravitransf}, assuming that $g\to \infty$ while keeping the coupling constant $\widehat g$ finite, yields the potential\footnote{The subscript ``flat'' refers to the Ricci-flat nature of the hypermultiplet target space -- not to the spacetime.}
\begin{equation}
\boxed{
\label{potential.zoomin1}
\mathscr{V}_\text{flat}=\frac{\widehat g^2\tilde\mu^2}{k^2}\,\left(\frac{1}{\widehat V}-3\widehat Z\right)
\,,}
\end{equation}
in the presence of a non-vanishing cosmological constant
\begin{equation}
\Lambda  =  -{3\widehat g^2 \over 2k^2}\,.
\end{equation}

Before proceeding with the alternative rigid limit, we would like to pause and make contact with the general results of Butter and Kuzenko on ${\cal N}=2$ supersymmetric $\sigma$-models in $\text{AdS}_4$ \cite{Butter:2011zt,Butter:2011kf}. According to these authors, rigid ${\cal N}=2$ supersymmetry in $\text{AdS}_4$ spacetime requires
the target space of the $\sigma$-model be 
a {\it non-compact} hyper-K\"ahler manifold endowed with a simply holomorphic isometry. Let $\xi$ be the simply holomorphic Killing field, and assume a basis for the complex structures obeying \eqref{rot}, so that the preserved one is $J_3$. The latter provides a {\it globally defined} Killing potential $\mathcal{K}$ as in \eqref{Killing.potential}. Following Butter and Kuzenko, the scalar potential reads:
\begin{equation}
\label{potentialflat.gen}
\mathscr{V}_{\text{BK}} =\mu^2\tilde\mu^2\left(\frac{1}{2}\, g_{\text{HK}}^{uv}\, \partial_u{\cal K}\, \partial_v{\cal K} 
- 3 \,{\cal K}\right),\quad \Lambda=-3\mu^2\,.
\end{equation}
Consider a generic hyper-K\"ahler metric as in  Eqs.~\eqref{generalHK} and  \eqref{BFmetric}, with a rotational isometry $\xi=\partial_\tau$ and Killing potential \eqref{Kcal.gen}. Inserting the latter into \eqref{potentialflat.gen}, we recover \eqref{potential.zoomin1} with $\mu=\frac{\widehat g}{\sqrt{2}\,k}$.

\paragraph{From Przanowski--Tod to Gibbons--Hawking} We now turn to the triple-scaling limit \eqref{zoomin1}, for which the Killing field becomes translational in the hyper-K\"ahler space. Performed on the scalar potential \eqref{potentialgravitransf}, while keeping $\widehat g$ finite, this limit yields
\begin{equation}
\boxed{
\label{potential.zoomin2}
\mathscr{V}_\text{flat}=\frac{\widehat g^2\tilde\mu^2}{k^2\,\widehat V}
\,.}
\end{equation}
The Killing $\partial_\tau$ is translational in the limit at hand, $k_-^2$ vanishes and so does $\Lambda$.  
The result  \eqref{potential.zoomin2} is in agreement with the potential of a hypermultiplet 
of global ${\cal N}=2$ supersymmetry in Minkowski spacetime \cite{Hull:1985pq}.
Along the same lines of thought, the alternative decoupling limit \eqref{alter.condit1} on \eqref{potentialgraviPT} yields 
\begin{equation}
\boxed{
\label{potential.zoomin3}
\mathscr{V}_\text{flat}=\frac{\widehat g^2\tilde\mu^2}{k^2\,Z_0^2\,\widehat V}\,,}
\end{equation}
where $\widehat g=\frac{g}{(k\tilde\mu)^{4/3}}$ is a finite coupling constant when $k\to0$.

\boldmath
\section{Spaces with Heisenberg isometry} \label{genquater}
\unboldmath

\subsection{The kinetic term and its three distinct rigid limits}\label{seclimits}

As already mentioned in the introduction, Heisenberg symmetry has a distinguished role. Firstly, only two hyper-K\"ahler spaces exist with this isometry group \cite{BEPS}, and one quaternion-K\"ahler \cite{Antoniadis:2015egy}. Secondly, the latter captures the type-II string one-loop perturbative corrections to the hypermultiplet scalar manifold  \cite{oneloop}, and can be derived through the quaternionic quotient method by gauging the $\mathscr{Z}$ isometry
inside the $Sp(2, 4)$ of the ${\cal N} = 2$ hypermultiplet manifold \cite{AADT}. The 
quaternion-K\"ahler space under consideration has actually extended $\text{Heisenberg}\ltimes U(1)$ isometry and its metric reads:
\begin{equation}
\begin{split}
\label{CPM}
&\mathrm{d}s_{\text{QK}}^2=\frac{8\rho^2}{V_1V_2^2}\left(\mathrm{d}\tau+\eta\mathrm{d}\varphi\right)^2+
\frac{2V_1}{V^2_2}\left(\mathrm{d}\rho^2+\mathrm{d}\eta^2+\mathrm{d}\varphi^2\right),\\
&V_1=\rho^2+2\sigma\geqslant0\,,\quad V_2=\rho^2-2\sigma\,,\quad \sigma=\text{constant}.
\end{split}
\end{equation}
The  $\text{Heisenberg}\ltimes U(1)$ algebra is realized as
\begin{equation}
\label{Heisenbquat}
\begin{split}
&[\mathscr{X},\mathscr{Y}]=\mathscr{Z}\,,\quad
 [\mathscr{M},\mathscr{X}]=\mathscr{Y}\,,\quad [\mathscr{M},\mathscr{Y}]=-\mathscr{X}\,,\\
&\mathscr{X}=\partial_\eta-\varphi\,\partial_\tau\,,\quad\mathscr{Y}=\partial_\varphi\,,
\quad\mathscr{Z}=\partial_\tau\,,\quad
\mathscr{M}=\varphi\,\partial_\eta-\eta\,\partial_\varphi+\frac12\,\left(\eta^2-\varphi^2\right)\partial_\tau\,.
\end{split}
\end{equation} 
For vanishing $\sigma$, this isometry algebra is actually extended to $U(1,2)$, and the quaternionic space becomes the non-compact 
$\nicefrac{SU(2,1)}{SU(2)\times U(1)}$ (see App. \ref{PFSmetric}).
In the frame at hand, the space \eqref{CPM} is fibered over $\mathscr{Z}$.

It was noticed in \cite{AADT} that the rigid limit of the kinetic term \eqref{kinetic} with \eqref{CPM} is $\sigma$-dependent.
For $\sigma\geqslant0$ two trivial, flat-space rigid limits occur around $\rho^2\sim0$ and $\rho^2-2\sigma\sim0$. 
For $\sigma<0$ a non-trivial limit appears at $\rho^2+2\sigma\sim0$. The latter case corresponds precisely to the hyper-K\"ahler metric \eqref{jdkdkfjh}, invariant under the
$\text{Heisenberg}\ltimes U(1)$ symmetry \eqref{Heisenbhyper}. These results can be summarized as follows:
\begin{description}
\item[-]\textsc{Parametric region $\sigma\geqslant0$}
\begin{enumerate}
\item
Around $\rho^2\sim0$ we zoom as:
\begin{equation}
\label{jopdjkf}
\rho=\sqrt{\sigma}\,k \tilde\mu\,t\,,\quad
\eta=\sqrt{\sigma}\,k \tilde\mu\,x\,,\quad
\varphi=\sqrt{\sigma}\,k \tilde\mu\,y\,,\quad
\tau=\sigma\, z\,,\quad k\to0\,,
\end{equation}
and this leads to flat space 
\begin{equation}
\label{jkfdlklfes}
\text{d}s_{\text HK}^2=\text{d}x^2+\text{d}y^2+\text{d}t^2+t^2\text{d}z^2\,,
\end{equation}
where  the Killing supporting the fiber in the original quaternionic space, $\mathscr{Z}\propto\partial_z$ is here a simply holomorphic (rotational) Killing vector.
\item
Around $\rho^2-2\sigma\sim0$ we zoom as:
\begin{equation}
\begin{split}
\label{jopdjkfaa}
&\rho^2=2\sigma+k \tilde\mu\left(1+k \tilde\mu\, t\right),\quad
\eta=\frac{1}{\sqrt{8\sigma}}\left(k \tilde\mu\right)^2x\,,\quad\\
&\varphi=\frac{1}{\sqrt{8\sigma}}\left(k \tilde\mu\right)^2y\,,\quad
\tau=\frac{1}{2}\,\left(k \tilde\mu\right)^2\,z\,,\quad k\to0\,,
\end{split}
\end{equation}
and obtain again flat space
\begin{equation}
\label{jkfdlklfesa}
\text{d}s_{\text HK}^2=\text{d}x^2+\text{d}y^2+\text{d}t^2+\text{d}z^2\,,
\end{equation}
where $\mathscr{Z}\propto\partial_z$ is now a triholomorphic (translational) Killing vector.
\end{enumerate}
\item[-]\textsc{Parametric region $\sigma<0$}

Around $\rho^2+2\sigma\sim0$ we zoom as:
\begin{equation}
\begin{split}
\label{limitHK}
&\rho^2=-2\sigma\left(1+2\,t (k\tilde\mu)^{\nicefrac{2}{3}}\right),
\quad \eta=\sqrt{-2\sigma} x (k\tilde\mu)^{\nicefrac{2}{3}}\,,\\
&\varphi=\sqrt{-2\sigma} y (k\tilde\mu)^{\nicefrac{2}{3}}\,,
\quad \tau=-2\sigma\, z (k\tilde\mu)^{\nicefrac{4}{3}}\,,\quad k\to0\,,
\end{split}
\end{equation}
such that the Kretschmann scalar of \eqref{CPM},
\begin{equation}
R_{xuvw}R^{xuvw}=48\frac{(\rho^4+4\sigma^2)(\rho^8+56\rho^4\sigma^2+16\sigma^4)}{(\rho^2+2\sigma)^6}\,,
\end{equation}
remains finite when $k\to0$. This limiting procedure results to the hyper-K\"ahler space \eqref{jdkdkfjh}, 
\begin{equation}
\label{jdkdkfjh2}
\text{d}s_{\text HK}^2 =\frac{1}{t} (\text{\text{d}}z + x \, \text{\text{d}}y)^2 + 
t\left(\text{\text{d}}t^2 + \text{\text{d}}x^2 +\text{\text{d}}y^2\right),
\end{equation}
which is invariant under the
$\text{Heisenberg}\ltimes U(1)$ symmetry \eqref{Heisenbhyper}, \eqref{Heisenbhyper1}. 
Therefore the decoupling limit \eqref{limitHK} 
preserves all the isometries of the quaternionic ancestor metric \eqref{CPM}. Along the process, the Killing supporting the fiber, $\mathscr{Z}\propto\partial_z$ becomes a triholomorphic (translational) Killing vector.

\end{description}

In order to make contact with the general developments presented in Sec. \ref{kineticrigid}, we should recast the quaternionic $\text{Heisenberg}\ltimes U(1)$-invariant metric \eqref{CPM} in the Przanowski--Tod  form \eqref{prztod1}. This is achieved by keeping $\tau$ unaltered, while trading $\rho, \eta, \varphi$ for $X,Y,Z$ as follows:
\begin{equation} 
X=\eta\,,\quad
Y=\varphi\,,
\quad Z=\frac{V_2(\rho)}{2}\,,
\end{equation}
and setting
\begin{equation} 
\omega=\eta\,\text{d}\varphi\,,
\quad V=\frac{V_1(\rho)}{2\rho^2}\,,
\quad \text{e}^\Psi=\rho^2\,.
\end{equation}
The fiber of the Przanowski--Tod is supported by the Killing field $\mathscr{Z}=\partial_\tau$, which turns rotational or translational in the hyper-K\"ahler, depending on the decoupling limit.

With these conventions, on the one hand,
the rigid limits for $\sigma\geqslant0$, 
$\rho^2\sim0$ and  $\rho^2-2\sigma\sim0$, correspond to \eqref{zoomin} (Przanowski--Tod to Boyer--Finley \emph{i.e.} hyper-K\"ahler limit with rotational Killing $\mathscr{Z}$) and  \eqref{zoomin1} (Przanowski--Tod to Gibbons--Hawking \emph{i.e.} hyper-K\"ahler limit with translational Killing $\mathscr{Z}$). On the other hand,
the rigid limit $\rho^2+2\sigma\sim0$ for $\sigma<0$ is the alternative Przanowski--Tod to Gibbons--Hawking limit, \eqref{alter.condit1} and \eqref{alter.condit2},  leading to the unique hyper-K\"ahler space with $\text{Heisenberg}\ltimes U(1)$ symmetry, with a translational Killing vector $\mathscr{Z}$ supporting the fiber.  

Notice finally that in all cases, extra Killing fields survive the decoupling limit, either simply holomorphic or 
triholomorphic, which may be chosen to further recast the hyper-K\"ahler metric in another  Boyer--Finley 
or Gibbons--Hawking frame. Such options can be exploited depending on the form of the original scalar 
potential, and its structure in the decoupling limit. 

\subsection{The scalar potential}

The form of the scalar potential depends on which symmetry is gauged, \emph{i.e.} which element 
is chosen inside the $\text{Heisenberg}\ltimes U(1)$ isometry group of the quaternionic metric \eqref{CPM}. We will  
always use the graviphoton for the gauging,
as already advertised, and the general formalism of Sec. \ref{gen.scalar.potential.cs}. 

In our analysis, we will systematically investigate the rigid limits. There are always three distinct cases 
corresponding to the limits
\eqref{jopdjkf}, \eqref{jopdjkfaa} and \eqref{limitHK},
exhibited for the kinetic term \eqref{kinetic} on the  $\text{Heisenberg}\ltimes U(1)$-invariant 
quaternionic space \eqref{CPM}, and leading to the hyper-K\"ahler spaces
\eqref{jkfdlklfes}, \eqref{jkfdlklfesa} and \eqref{jdkdkfjh2}. They are associated with the 
Przanowski--Tod to Boyer--Finley limit (hyper-K\"ahler limit with rotational Killing), the  Przanowski--Tod to 
Gibbons--Hawking limit (hyper-K\"ahler limit with translational Killing), and the Przanowski--Tod to 
Gibbons--Hawking limit with full $\text{Heisenberg}\ltimes U(1)$ symmetry. 

Coming back to the potential term, two distinct situations arise depending on which isometry is
 gauged: (\romannumeral1) either the field carrying the Przanowski--Tod fiber $\mathscr{Z}$
-- the potential is $\text{Heisenberg}\ltimes U(1)$ invariant, 
(\romannumeral2) or any other Killing -- the potential is
not $\text{Heisenberg}\ltimes U(1)$ invariant. In the first instance, the properties of the potential 
and its associated spectrum at the supergravity level and in the decoupling limits will 
follow the general classification presented in Sec. \ref{gen.scalar.potential}. 
For these limits, in particular, we will find (Sec. \ref{gen.scalar.potential.dc}) 
rigid $\mathcal{N}=2$ supersymmetry with potential \eqref{potential.zoomin1} and 
anti-de Sitter vacuum, or with potentials \eqref{potential.zoomin2}, \eqref{potential.zoomin3} 
and Minkowski vacuum. The case (\romannumeral2) is expected to be slightly different, 
and is worth presenting case by case: $\mathscr{M}$, $\mathscr{Z}$ plus $\mathscr{M}$, $\mathscr{Y}$. 
In the three rigid-supersymmetry limits, we find only anti-de Sitter vacuums for the first, 
and anti-de Sitter and Minkowski for the second and the third. The $\mathscr{Y}$ gauging 
exhibits a further peculiarity already at the supergravity level, namely a de Sitter 
$\mathcal{N}=0$ vacuum, which stands outside of the general analysis of Sec. \ref{gen.scalar.potential.cs}, 
where supersymmetry was unbroken.

\paragraph{Gauging the $\mathscr{Z}$ isometry}

Here, the Killing vector supporting the gauging is $\xi=g\,\mathscr{Z}$ (see \eqref{Heisenbquat}). 
The scalar potential is obtained thanks to the general formula \eqref{potentialgravi}: 
\begin{equation}
\label{lmoskp}
\mathscr{V}_\xi=\frac{2g^2}{k^4}\,\frac{\rho^2-6\sigma}{V_1V_2^2}\,,
\end{equation}
which is invariant under the action of $\text{Heisenberg}\ltimes U(1)$.
The potential at hand has an extremum at the origin $\rho=0$, the fixed point of $\xi$, and 
two flat directions $\eta,\varphi$. 
In order to analyze this extremum, we move from polar to Cartesian coordinates
\begin{equation}
\label{polarCartesian}
(\rho,\tau)\mapsto (q_1,q_2):\quad \rho=\sqrt{\sigma\left(q_1^2+q_2^2\right)}\,,\quad \tau=\sigma\arctan\frac{q_2}{q_1}\,,
\end{equation}
as $\rho=0$ is a coordinate singularity of \eqref{CPM}.

Expanding \eqref{lmoskp} at second order  around the extremum:
\begin{equation}
\label{jkfklldlld}
\mathscr{V}_\xi\approx\frac{\widehat g^2}{k^4}\,\left(-\frac{3}{2}-\frac{q_1^2+q_2^2}{2}\right)\,,
\end{equation}
where $\widehat g=g/\sigma$. From the latter, and normalizing with respect to the kinetic term, Eqs.~\eqref{kinetic} and \eqref{CPM}, we read off the mass terms 
and the cosmological constant: 
\begin{equation}
\label{}
M_{q_1,q_2}^2=-\frac{\widehat g^2}{k^2}=\frac{2}{3}\Lambda\,,\quad M_{\eta,\varphi}^2=0\,,\quad\Lambda= -{3\widehat g^2 \over 2k^2}\,.
\end{equation}
These satisfy the Breitenlohner--Freedman stability bound, and fit the general form \eqref{genmasspec} of the mass spectrum of an ${\cal N}=2$ chiral multiplet 
in $\text{AdS}_4$ spacetime, 
with $A=\{q_1,q_2\}$, $B=\{\eta,\varphi\}$ and
$\mu=m=\frac{\widehat g}{\sqrt{2}\, k}$. Besides the flat directions $\eta,\varphi$, 
the fields $q_{1,2}$ are massless.

Next task in our agenda is to analyze the behaviour of the potential \eqref{lmoskp} in the three distinguished rigid limits
\eqref{jopdjkf}, \eqref{jopdjkfaa} and \eqref{limitHK}.
\begin{description}
 \item[-] 
\textsc{Parametric region $\sigma\geqslant0$}

\begin{enumerate}
\item

Applying the rigid limit \eqref{jopdjkf} to \eqref{lmoskp}, we find the potential 
\begin{equation}
\label{potentialflate}
\mathscr{V}_{\text{flat}}=-\frac{\widehat g^2\tilde\mu^2}{2k^2}\,t^2\,.
\end{equation}
To this end, we use the generic expression for the potential in the Przanowski--Tod to Boyer--Finley rigid limit, Eq.~\eqref{potential.zoomin1}, after rewriting the 
flat space \eqref{jkfdlklfes} in the Boyer--Finley frame along the rotational isometry $\mathscr{Z}=\partial_\tau=\frac{1}{\sigma}\partial_z$, for which
\begin{equation}
\frac{1}{\widehat V}=2 \widehat Z= t^2\,.
\end{equation}

In the present rigid limit, the vacuum is thus an $\text{AdS}_4$ spacetime, and the hyper-K\"ahler space describes a 
global ${\cal N}=2$ hypermultiplet.

\item The rigid limit under consideration is now \eqref{jopdjkfaa}. Applied to \eqref{lmoskp}, it leads to 
 a global ${\cal N}=2$ hypermultiplet in Minkowski spacetime with an irrelevant constant potential
\begin{equation}
\mathscr{V}_{\text{flat}}=-\frac{\widetilde g^2\tilde\mu^2}{2k^2}\,,
\end{equation}
where $\widetilde g=\frac{2\sigma\, \widehat g}{(k\tilde\mu)^2}$  is a finite coupling constant when $k\to0$.

\end{enumerate}

\item[-] 
\textsc{Parametric region $\sigma<0$}

The scalar potential \eqref{lmoskp} becomes
\begin{equation}
\mathscr{V}_{\text{flat}}=\frac{\widetilde g^2\tilde\mu^2}{k^2\,t}\,,
\end{equation} 
in the rigid limit \eqref{limitHK}, with $\widetilde g=-\frac{\widehat g}{2(k\tilde\mu)^{4/3}}$ a finite coupling constant at $k\to0$.
Again, the cosmological constant vanishes and we describe a global ${\cal N}=2$ hypermultiplet in Minkowski spacetime.
\end{description}

\paragraph{Gauging the $\mathscr{M}$ isometry}

Consider now the isometry of \eqref{CPM} generated by $\xi=g\,\mathscr{M}$. Using \eqref{potentialgravi} 
we find the following scalar potential:
\begin{equation}
\label{jjfkldfl}
\mathscr{V}_\xi=\frac{g^2}{k^4}\,\left(-\frac{3}{2}-\frac{V_1}{V_2^2}\left(\eta^2+\varphi^2\right)+
\frac{V_2-4\sigma}{2V_1V_2^2}\left(\eta^2+\varphi^2\right)^2\right),
\end{equation}
invariant under the action of $\mathscr{M}$ and $\mathscr{Z}$. This potential has an extremum at the fixed point of $\xi$, $\eta=\varphi=0$, and two flat directions $\rho,\tau$. We can directly read off the mass terms and the cosmological constant (again normalization with respect to the kinetic term is required):  
\begin{equation}
M_{\rho,\tau}^2=0\,,\quad M_{\eta,\varphi}^2=-\frac{g^2}{k^2}=\frac{2}{3}\Lambda\,,\quad\Lambda= -{3g^2\over 2k^2}\,.
\end{equation}
Comparing with the general expression \eqref{genmasspec}, we identify the fields  $A=\{\eta,\varphi\}$ and 
$B=\{\rho,\tau\}$, whereas $\mu=m=\frac{g}{\sqrt{2}\, k}$. We find again the spectrum of an 
${\cal N}=2$ chiral multiplet in $\text{AdS}_4$, with two massless fields $\rho,\tau$ and two 
 massive fields $\eta,\varphi$.

Let us now consider the usual rigid limits when the dynamics is captured by the potential \eqref{jjfkldfl}.
\begin{description}
\item[-]
\textsc{Parametric region $\sigma\geqslant0$}

The pattern goes as in the previous gauging.  We first consider the rigid limits 
\eqref{jopdjkf} or \eqref{jopdjkfaa}.
We use again Eq.  \eqref{potential.zoomin1}
after rewriting the flat space \eqref{jkfdlklfes} or \eqref{jkfdlklfesa} in the Boyer--Finley frame along the rotational isometry $\mathscr{M}$, with
\begin{equation}
\frac{1}{\widehat V}=x^2+y^2\,,\quad \widehat Z=\frac12\,\left(x^2+y^2\right).
\end{equation}
We thus find for both limits $\text{AdS}_4$ spacetime 
with $\Lambda  =  -{3g^2 \over 2k^2}$. The hyper-K\"ahler space describes a 
a global ${\cal N}=2$ hypermultiplet with potential
\begin{equation}
\mathscr{V}_{\text{flat}}=-\frac{g^2\tilde\mu^2}{2k^2}\,\left(x^2+y^2\right).
\end{equation}
 \item[-] 
\textsc{Parametric region $\sigma<0$} 

Applying the rigid limit \eqref{limitHK} to \eqref{jjfkldfl},  we utilize Eq.~\eqref{potential.zoomin1} after
rewriting \eqref{jdkdkfjh} in the Boyer--Finley frame 
along the rotational isometry $\mathscr{M}$ \cite{Antoniadis:2015egy}: 
\begin{equation}
\label{jflmflaosomd}
 \frac{1}{\widehat V}=\frac{1}{4t}(x^2+y^2)^2+t(x^2+y^2)\,,\quad  \widehat Z=\frac{1}{2}t(x^2+y^2)\,.
\end{equation}
We now find a 
global ${\cal N}=2$ hypermultiplet in $\text{AdS}_4$ spacetime with
with potential
\begin{equation}
\label{potentialflat1}
\mathscr{V}_{\text{flat}}=-\frac{g^2\tilde\mu^2}{2k^2}\,\left(t(x^2+y^2)
-\frac{1}{2t}\,\left(x^2+y^2\right)^2\right),
\end{equation}
and cosmological constant 
$
\Lambda  =  -{3g^2 \over 2k^2}$. 

A comment is worth here on the Killing potential ${\cal K}$. This is determined 
using \eqref{Kcal.gen} and \eqref{jflmflaosomd}:
\begin{equation}
{\cal K}=2\widehat Z=t (x^2+y^2)\,.
\end{equation}
In the K\"ahler coordinates \eqref{KahlerM1}, the Killing potential ${\cal K}$ reads:
\begin{equation}
{\cal K}=\frac12\,K+W+\overline W\,,\quad W=-\frac16\,\Phi^3\,.
\end{equation}
Hence ${\cal K}$ appears as a K\"ahler potential and $W$ is
a K\"ahler transformation. 
Notice also that in the set of K\"ahler coordinates \eqref{KahlerM1},
the complex structure $J_1$ is diagonalized while $\mathscr{M}$ is not holomorphic.

\end{description}

\paragraph{Gauging  $\mathscr{Z}$ and $\mathscr{M}$ isometries}
This gauging is performed along  $\xi=g_1\mathscr{Z}+g_2\mathscr{M}$. The scalar potential is determined with  \eqref{potentialgravi} and \eqref{CPM}, and is invariant under the action of $\mathscr{M}$ and $\mathscr{Z}$: 
\begin{eqnarray}
\mathscr{V}_\xi&=&\frac{2g_1^2}{k^4}\,\frac{\rho^2-6\sigma}{V_1V_2^2}+
\frac{g_2^2}{k^4}\,\left(-\frac{3}{2}-\frac{V_1}{V_2^2}\left(\eta^2+\varphi^2\right)+
\frac{V_2-4\sigma}{2V_1V_2^2}\left(\eta^2+\varphi^2\right)^2\right)\nonumber\\
&&-2g_1g_2\frac{\left(\eta^2+\varphi^2\right)\left(V_2-4\sigma\right)+3V_1V_2}{k^4V_1V_2^2}\,.
\label{jjfkldfsdsl}
\end{eqnarray}
This potential has an extremum at $\rho=\eta=\varphi=0$, the fixed point of $\xi$. 
To analyze this extremum, we change coordinates from polar to Cartesian as in \eqref{polarCartesian},
and we expand \eqref{jjfkldfsdsl} at second-order around the extremum 
\begin{equation}
\label{jjfkldfsdsler}
\mathscr{V}_\xi\approx\frac{1}{k^4}\,\left(-\frac{3(\widehat g_1-g_2)^2}{2}+\frac12\,\widehat g_1(3g_2-\widehat g_1)(q_1^2+q_2^2)+
\frac{g_2}{2\sigma}\,(3\widehat g_1-g_2)(\eta^2+\varphi^2)\right),
\end{equation}
where $\nicefrac{\widehat g_1=g_1}{\sigma}$. 
The mass terms and the cosmological constant are obtained as usual: 
\begin{equation}
M_{q_1,q_2}^2=\frac{g_1(3g_2-\widehat g_1)}{k^2}\,,\quad M_{\eta,\varphi}^2=\frac{g_2(3\widehat g_1-g_2)}{k^2}\,,
\quad\Lambda= -{3(\widehat g_1-g_2)^2 \over 2k^2}\,,
\end{equation}
and satisfy the Breitenlohner--Freedman stability bound. For $\widehat g_1\neq g_2$, this vacuum generically describes the spectrum of a hypermultiplet in $\text{AdS}_4$ and 
fits the general form \eqref{genmasspec} of the mass spectrum 
with $A=\{\eta,\varphi\}$, $B=\{q_1,q_2\}$ and
\begin{equation}
\mu=\frac{g_2-\widehat g_1}{\sqrt{2}\,k}\,,\quad 
m=\frac{\widehat g_1+g_2}{\sqrt{2}\,k}\,.
\end{equation}

We now come to consider the rigid limits of the kinetic term \eqref{kinetic}, \eqref{CPM} on the potential \eqref{jjfkldfsdsl}.
\begin{description}
 \item[-] 
\textsc{Parametric region $\sigma\geqslant0$}

\begin{enumerate}

\item

Applying the rigid limit \eqref{jopdjkf} to \eqref{jjfkldfsdsl}, we obtain:
\begin{equation}
\mathscr{V}_\xi=-\frac{3\left(\widehat g_1-g_2\right)^2}{2k^4}-
\frac{\tilde\mu^2}{2k^2}\,\left(\widehat g_1^2t^2+g_2^2(x^2+y^2)-3\widehat g_1g_2(t^2+x^2+y^2)\right).
\end{equation}
Two distinct cases should be examined, when  $\widehat g_1\neq g_2$ or $\widehat g_1=g_2$\,:\begin{enumerate}
\item
For $\widehat g_1\neq g_2$,
we find a global ${\cal N}=2$ hypermultiplet in $\text{AdS}_4$ spacetime with $\Lambda  = -\frac{3\left(\widehat g_1-g_2\right)^2}{2k^2}$
and potential
\begin{equation}
\label{potentialflate134}
\mathscr{V}_{\text{flat}}=-
\frac{\tilde\mu^2}{2k^2}\,\left(\widehat g_1^2t^2+g_2^2(x^2+y^2)-3\widehat g_1g_2(t^2+x^2+y^2)\right).
\end{equation}
To prove this, we utilize \eqref{potential.zoomin1} after rewriting
 the hyper-K\"ahler metric \eqref{jkfdlklfes} in the Boyer--Finley frame along the rotational isometry 
 $\xi=\widehat g_1\mathscr{Z}+g_2\mathscr{M}$. We find
\begin{equation*}
 \frac{1}{\widehat V}=\widehat g_1^2t^2+g_2^2\left(x^2+y^2\right),\quad  
 \widehat Z=\frac12\,\left(\widehat g_1-g_2\right)\left(\widehat g_1t^2-g_2\left(x^2+y^2\right)\right).
\end{equation*}

\item
For $\widehat g_1= g_2$, we obtain the potential:\footnote{
For $\widehat g_1=g_2$, the isometry is translational, in agreement with a theorem in 
\cite{GibbonsRubback} on commuting rotational isometries,
and the rigid limit yields
a global ${\cal N}=2$ hypermultiplet in Minkowski spacetime.
}

\begin{equation}
\mathscr{V}_{\text{flat}}=\frac{\tilde\mu^2\widehat g_1^2}{k^2}\,\left(t^2+x^2+y^2\right)\,,
\end{equation}
which corresponds to a massive ${\cal N}=2$ hypermultiplet in Minkowski spacetime.
Indeed this can be easily seen by changing coordinates in \eqref{jkfdlklfes}
from polar $(t,z)$ to Cartesian ones $(q_1,q_2)$, as $t=0$ is a coordinate singularity of the
hyper-K\"ahler metric \eqref{jkfdlklfes}.

\end{enumerate}

\item
We now consider the rigid limit \eqref{jopdjkfaa} to \eqref{jjfkldfsdsl}. We rewrite the hyper-K\"ahler metric \eqref{jkfdlklfesa} in the Boyer--Finley frame along the rotational isometry 
 $\xi=\widetilde g_1\mathscr{Z}+g_2\mathscr{M}$ and use  \eqref{potential.zoomin1}: 
 \begin{equation}
 \frac{1}{\widehat V}=\widetilde g_1^2+g_2^2\left(x^2+y^2\right),\quad  
 \widehat Z=\frac12\,\left(\widetilde g_1^2+g_2^2(x^2+y^2)-2\widetilde g_1 g_2 t\right).
\end{equation}
We obtain a 
global ${\cal N}=2$ hypermultiplet in $\text{AdS}_4$ spacetime with potential
\begin{equation}
\mathscr{V}_{\text{flat}}=-
\frac{\tilde\mu^2}{2k^2}\,\left(\widetilde g_1^2-6\widetilde g_1 g_2 t+g_2^2(x^2+y^2)\right),
\end{equation}
and $
\Lambda  =  -{3g_2^2 \over 2k^2}$, where $\widetilde g_1=\frac{2\sigma\,\widehat g_1}{(k\tilde\mu)^2}$  is a finite coupling constant when $k\to0$.

\end{enumerate}

 \item[-] 
\textsc{Parametric region $\sigma<0$}

The last rigid limit is  \eqref{limitHK} applied to \eqref{jjfkldfsdsl}. Now, with \eqref{potential.zoomin1}, we express the 
hyper-K\"ahler metric \eqref{jdkdkfjh2} in the Boyer--Finley frame along the rotational isometry 
 $\xi=\widetilde g_1\mathscr{Z}+g_2\mathscr{M}$:
 \begin{equation}
 \begin{split}
& \frac{1}{\widehat V}=\frac{4\widetilde g_1^2-4\widetilde g_1 g_2(x^2+y^2)+g_2^2(x^2+y^2)(4t^2+x^2+y^2)}{4t}\,,\\
 &\widehat Z=\frac12\,g_2t\left(-2\widetilde g_1+g_2(x^2+y^2)\right).
 \end{split}
\end{equation}
Again, we find a 
global ${\cal N}=2$ hypermultiplet in $\text{AdS}_4$ spacetime with $\Lambda  =  -{3g_2^2 \over 2k^2}$ and
potential
\begin{equation}
\mathscr{V}_{\text{flat}}=\frac{\tilde\mu^2}{4tk^2}\,\left(4\widetilde g_1^2+4\widetilde g_1 g_2(3t^2-x^2-y^2)-g_2^2(2t^2-x^2-y^2)(x^2+y^2)\right),
\end{equation}
where $\widetilde g_1=-\frac{\widehat g_1}{2(k\tilde\mu)^{4/3}}$ is a finite coupling constant when $k\to0$.

\end{description}

\paragraph{Gauging the $\mathscr{Y}$ isometry} This last gauging is based on the isometry  $\xi=g\,\mathscr{Y}$ and leads to the scalar potential
\begin{equation}
\label{jjfkldfsdslhfjfk}
\mathscr{V}_\xi=\frac{2g^2}{k^4}\,\frac{\eta^2(V_2-4\sigma)-2V_1(V_2+\sigma)}{V_1V_2^2}\, ,
\end{equation}
invariant under the action of $\mathscr{Y}$ and $\mathscr{Z}$. 

The potential \eqref{jjfkldfsdslhfjfk} has an extremum at $\rho=\eta=0$, and a flat direction $\varphi$. 
As usual, we change coordinates from polar to Cartesian, see Eqs. \eqref{polarCartesian},
and we expand \eqref{jjfkldfsdslhfjfk} at second order 
\begin{equation}
\mathscr{V}_\xi\approx\frac{g^2}{\sigma k^4}-\frac{3 g^2\eta^2}{2\sigma^2k^4}\,,
\end{equation}
where $\sigma>0$, required by  positivity of the metric \eqref{CPM} around the extremum at hand.
Using the latter  and normalizing with \eqref{CPM}, we read off the mass terms and the cosmological constant:
\begin{equation}
M_\eta^2=-\frac{3 g^2}{\sigma k^2}\,,\quad M_\varphi^2=M_{q_1}^2=M_{q_2}^2=0\,,\quad \Lambda=\frac{g^2}{\sigma k^2}\,.
\end{equation}
As advertised earlier on, this corresponds 
to an ${\cal N}=0$ hypermultiplet in $\text{dS}_4$ spacetime with ${\cal R}=4\Lambda$.

Our next task is the analysis of the usual rigid limits on the potential \eqref{jjfkldfsdslhfjfk}.
\begin{description}
 \item[-] 
\textsc{Parametric region $\sigma\geqslant0$}

\begin{enumerate}

\item
The rigid limit \eqref{jopdjkf} applied to \eqref{jjfkldfsdslhfjfk} leads to a 
global ${\cal N}=2$ hypermultiplet in Minkowski spacetime with constant potential
$
\mathscr{V}_{\text{flat}}=\frac{\widehat g^2\tilde\mu^2}{k^2}$,
where $\widehat g=\frac{g}{\sqrt{\sigma}k\tilde\mu}$  is a finite coupling constant when $k\to0$.

\item
In the alternative rigid limit \eqref{jopdjkfaa}, we find again a 
global ${\cal N}=2$ hypermultiplet in Minkowski spacetime (with another constant potential
$
\mathscr{V}_{\text{flat}}=-\frac{\widehat g^2\tilde\mu^2}{2k^2}\,,
$
and $\widehat g=\frac{\sqrt{8\sigma}\,  g}{(k\tilde\mu)^2}$  a coupling constant, finite in the decoupling).

\end{enumerate}

 \item[-] 
\textsc{Parametric region $\sigma<0$}

Similarly the rigid limit \eqref{limitHK} of the potential \eqref{jjfkldfsdslhfjfk} gives a
global ${\cal N}=2$ hypermultiplet in Minkowski spacetime with potential
\begin{equation}
\mathscr{V}_{\text{flat}}=\frac{\widehat g^2\tilde\mu^2(t^2+x^2)}{k^2t}\,,
\end{equation}
where $\widehat g=\frac{ g}{\sqrt{-2\sigma}(k\tilde\mu)^{4/3}}$ remains finite when $k\to0$.

It is worth mentioning, that this potential could be obtained from the ${\cal N}=1$ expression
\begin{equation}
\mathscr{V}=\frac{\tilde\mu^2}{k^2}\,K^{a\overline b} W_{a}\overline W_{\overline b}\,, 
\end{equation}
expressed in terms of a linear holomorphic superpotential: $W=\widehat g\, T$\,, 
in the K\"ahler basis \eqref{KahlerM1}.
This linear superpotential breaks supersymmetry as the kinetic term is non-canonical.

\end{description}

\section*{Conclusion}

We can now highlight and summarize our results. The core of the present work is the investigation of 
various off-shell gravity decoupling limits of the ${\cal N}=2$ scalar hypermultiplet. This concerns at the first place the kinetic term, based on a $\sigma$-model which has a quaternion-K\"ahler target space. Analyzing the decoupling limit,
establishes various relationships between quaternionic and hyper-K\"ahler spaces with symmetry.

Equally important is the behaviour of the scalar potential, produced by gauging symmetries with specific vectors. For the general analysis, we have chosen the graviphoton, along a generic isometry of a quaternion-K\"ahler space of the Przanowski--Tod type. The rigid limits reveal two separate 
cases: a rigid ${\cal N}=2$ theory on Minkowski or on $\text{AdS}_4$ spacetime, depending on 
whether the isometry is translational or rotational in the hyper-K\"ahler limit. These results are in agreement with previous results in the literature 
for global ${\cal N}=2$ in Minkowski and $\text{AdS}_4$ spaces \cite{AlvarezGaume:1981hm} and \cite{Butter:2011zt, Butter:2011kf}.

In order to illustrate our general results, we analyzed extensively  the quaternionic
metric with $\text{Heisenberg}\ltimes U(1)$ isometry, Eq.~\eqref{CPM}. The global ${\cal N}=2$ limits of this space 
are found to be trivial (flat space) or the hyper-K\"ahler space \eqref{jdkdkfjh2}, 
which is invariant under the
$\text{Heisenberg}\ltimes U(1)$ symmetry \eqref{Heisenbhyper}, \eqref{Heisenbhyper1}.
We further derived the scalar potential by
gauging the graviphoton along all possible isometries of the quaternion-K\"ahler space, $(\mathscr{Y},\mathscr{Z}, \mathscr{M})$ 
and studied the vacuum structure of the scalar potential together with its on/off-shell rigid limits. 

Interesting open questions remain at this stage, which have not been addressed in our work. An important one is to 
gauge isometries of the hypermultiplet $\sigma$-model using the graviphoton \emph{and} a vector multiplet.
The latter combination can give access to vacuum solutions which describe the spectrum of ${\cal N}=1$
hypermultiplets in Minkowski or $\text{AdS}_4$ spacetimes.

\addcontentsline{toc}{section}{Conclusion}

\section*{Acknowledgements}

We would like to thank Jean-Claude Jacot for participation in the early stages of this project.  
We also acknowledge each-other institutes for hospitality and financial support. 
 This work was partially supported by the \textsl{Germaine de Sta\"el} Franco--Swiss bilateral 
 program 2015 (project no 32753SG). K. Siampos would like to thank the 
 TH-Unit at CERN for hospitality and financial support during various stages of this project. 
He would like also to thank the ICTP, Trieste and the Physics department of the National and Kapodistrian 
University of Athens for hospitality.

\appendix

\boldmath
\section{Pseudo-Fubini--Study metric}
\label{PFSmetric}
\unboldmath

We shall utilize the results of Secs. \ref{kineticrigid} and \ref{gen.scalar.potential} for the
pseudo-Fubini--Study metric on $\widetilde{CP}_2=\nicefrac{SU(1,2)}{SU(2)\times U(1)}$
\begin{equation}
\label{FubiniQK}
\mathrm{d}s_{\text{QK}}^2=2g_{a\ov b}\,\mathrm{d}z^a\mathrm{d}\ov z^{\ov b}\,,\quad
g_{a\ov b}=K_{a\ov b}\,,\quad K=-\ln\left(1-|z|^2-|w|^2\right),
\quad z^a=(z,w)\,,
\end{equation}
This is a K\"ahler--Einstein--WSD space, with $R=-12$, 
which describes the universal hypermultiplet at string tree-level.

The K\"ahler potential is invariant under the action of
\begin{equation}
\label{CPiso}
\xi=\alpha\, i\left(z\,\partial_z-\ov z\,\partial_{\ov z}\right)+
\beta\, i\left(w\,\partial_w-\ov w\,\partial_{\ov w}\right),\quad \alpha,\beta\in\mathbb{R}\,,
\end{equation}
and through  Eq.~\eqref{potentialgravi} we find the scalar potential
\begin{equation}
\label{FubiniQKpot}
\mathscr{V}_\xi=\frac{(1-r_2^2)(4r_1^2+3r_2^2-3)\alpha^2+(1-r_1^2)(4r_2^2+3r_1^2-3)\beta^2+
2\alpha\beta(3-3r_1^2-3r_2^2+r_1^2r_2^2)}{2k^4(1-r_1^2-r_2^2)^2}\,,
\end{equation}
with $r_1^2=|z|^2$ and $r_2^2=|w|^2$. 

This potential has an extremum at $(z,w)=(0,0)$, fixed point of $\xi$.
Expanding it around this point we find at second order
\begin{equation}
\label{jfkfkff}
\mathscr{V}_\xi\approx\frac{1}{k^4}\,\left(-\frac{3}{2}\,(\alpha-\beta)^2+\alpha\left(3\beta-\alpha\right)r_1^2+
\beta\left(3\alpha-\beta\right)r_2^2\right)\,,
\end{equation}
and we read off the masses and the cosmological constant which satisfy the Breitenlohner--Freedman stability bound
\begin{equation}
M_{r_1}^2=\frac{\alpha\left(3\beta-\alpha\right)}{k^2}\,,\quad
M_{r_2}^2=\frac{\beta\left(3\alpha-\beta\right)}{k^2}\,,
\quad\Lambda=-\frac{3(\alpha-\beta)^2}{2k^2}\,.
\end{equation}
This vacuum generically describes the spectrum of a chiral multiplet
in an $\text{AdS}_4$ spacetime, similarly to \eqref{genmasspec}
\begin{equation}
M_{r_1}^2=m^2-2\mu^2-m\mu\,,\quad M_{r_2}^2=m^2-2\mu^2+m\mu\,,
\end{equation}
with $\Lambda=-3\mu^2$, where:
\begin{equation}
\mu=\frac{\alpha-\beta}{\sqrt{2}\,k}\,,\quad 
m=\frac{\alpha+\beta}{\sqrt{2}\,k}\,.
\end{equation}

We can now analyze the rigid limit around $z,w\sim0$. The kinetic term \eqref{kinetic} corresponding to \eqref{FubiniQK}, has a unique trivial gravity decoupling limit
around $z,w\sim0$, where we zoom as:
\begin{equation}
z=k\tilde\mu\,\widehat z\,,\quad w=k\tilde\mu\,\widehat w\,,\quad k\to0\,,
\end{equation}
leading to flat space
\begin{equation}
\label{hdkldmxsls}
\text{d}s_{\text HK}^2=2\left(\text{d}\widehat z\,\text{d}\widehat{\ov z}+\text{d}\widehat w\,\text{d}\widehat{\ov w}\right),
\end{equation}
while the potential \eqref{FubiniQKpot} truncates to:
\begin{equation}
\label{kododjfkf}
\mathscr{V}_\xi=-\frac{3(\alpha-\beta)^2}{2k^4}\,+\frac{\tilde\mu^2}{k^2}\,\left(\alpha\left(3\beta-\alpha\right)|\widehat z|^2+
\beta\left(3\alpha-\beta\right)|\widehat w|^2\right).
\end{equation}

Two cases need to be considered, for  $\alpha\neq\beta$ or $\alpha=\beta$\,:

\begin{enumerate}
\item
For $\alpha\neq\beta$,
we find a global ${\cal N}=2$ hypermultiplet in $\text{AdS}_4$ spacetime with potential
\begin{equation}
\label{potentialflate1}
\mathscr{V}_{\text{flat}}=\frac{\tilde\mu^2}{k^2}\,\left(\alpha\left(3\beta-\alpha\right)|\widehat z|^2+
\beta\left(3\alpha-\beta\right)|\widehat w|^2\right).
\end{equation}
and a non-vanishing cosmological constant
$
\Lambda  =  -{3(\alpha-\beta)^2 \over 2k^2}\,.
$
To prove this, we utilize \eqref{potential.zoomin1} after rewriting
 the hyper-K\"ahler metric \eqref{hdkldmxsls} in the Boyer--Finley frame along the rotational isometry \eqref{CPiso}
\begin{equation}
\label{CPHKBF}
 \frac{1}{\widehat V}=2\left(\alpha^2\,|\widehat z|^2+\beta^2\,|\widehat w|^2\right),\quad  
 \widehat Z=\left(\alpha-\beta\right)\left(\alpha\,|\widehat z|^2-\beta\,|\widehat w|^2\right).
\end{equation}

\item
For $\alpha=\beta$, we have  a global ${\cal N}=2$ hypermultiplet in Minkowski spacetime with potential:
\begin{equation}
\mathscr{V}_{\text{flat}}=\frac{2\tilde\mu^2}{k^2}\,\alpha^2\,
\left(|\widehat z|^2+|\widehat w|^2\right)\,,
\end{equation}
which corresponds to a massive ${\cal N}=2$ hypermultiplet in Minkowski spacetime.

\end{enumerate}

\boldmath
\section{Four-dimensional K\"ahler spaces with an isometry}
\label{Kahleriso}
\unboldmath

We are interested in providing an alternative exhibition of generic, Ricci-flat, scalar-flat or Einstein four-dimensional K\"ahler spaces with a holomorphic isometry. Hyper-K\"ahler (Ricci-flat), scalar-flat or Einstein solutions appear as Gibbons--Hawking like metrics. 
For generic and Ricci-flat four-dimensional K\"ahler space with a holomorphic isometry, see also Ref.~\cite{Chimento:2016run}.

\boldmath
\subsection{Four-dimensional K\"ahler spaces}
\unboldmath

We begin with K\"ahler complex coordinates $T,\Phi$, with an isometry acting as a shift of $\text{Im} T$ and
so the K\"ahler potential takes the form $K=K(T+\ov T,\Phi,\ov\Phi)$. A simple rearrangement yields
\begin{equation}
\label{genKahler}
\begin{split}
 \text{d}s&_{\text{K\"ahler}}^2=K_{T\ov T}\,\text{d}\,T\text{d}\ov T+K_{T\ov \Phi}\,\text{d}T\text{d}\ov \Phi+
 K_{\Phi\ov T}\,\text{d}\Phi\text{d}\ov T+K_{\Phi\ov \Phi}\,\text{d}\Phi\text{d}\ov \Phi\\
&=K_{T\ov T}\,\left(\text{d} \text{Im}\,T+\frac{i}{2K_{T\ov T}}\,
\left(K_{T\ov \Phi}\text{d}\ov\Phi-K_{\ov T\Phi}\text{d}\Phi \right)\right)^2+
\frac{1}{K_{T\ov T}}\left(\frac14\,\left(\text{d}K_T\right)^2+{\cal D}\,\text{d}\Phi\text{d}\ov\Phi\right),
\end{split}
\end{equation}
where ${\cal D}$ is the determinant of the K\"ahler metric
\begin{equation}
\label{detKahler}
{\cal D}:=\det K_{a\ov b}=K_{T\ov T}K_{\Phi\ov\Phi}-K_{T\ov\Phi}K_{\ov T\Phi}\,. 
\end{equation}
Next we define a new coordinate $Z$ as 
\begin{equation}
\label{Legendre0}
K_T:=\frac12\left(Z+c\right),
\end{equation}
where $c$ is an integration constant that can be absorbed by a K\"ahler transformation
\begin{equation*}
K\mapsto  K+\frac{c}{2}\,\left(T+\ov T\right).
\end{equation*}
We may think of \eqref{Legendre0} defining a Legendre transformation as
\begin{equation}
H(Z+c,\Phi,\ov\Phi)=\frac{1}{2}\,\left(Z+c\right)\left(T+\ov T\right)-K(T+\ov T,\Phi,\ov\Phi)\,,
\end{equation}
with transformation relations
\begin{equation}
\label{Legendre}
H_Z=\frac12\,\left(T+\ov T\right),\quad H_\Phi=-K_\Phi\,,\quad H_{\ov\Phi}=-K_{\ov\Phi}\,,
\end{equation}
which lead to\footnote{Using the Jacobian matrix for the Legendre transformation 
$\left(T+\ov T,\Phi,\ov\Phi\right)\mapsto \left(Z,\Phi,\ov\Phi\right),$ implied by \eqref{Legendre}.
}
\begin{equation}
K_{T\ov T}=\frac{1}{4H_{ZZ}}\,,\quad K_{T\ov\Phi}=-\frac{H_{Z\ov\Phi}}{2H_{ZZ}}\,,\quad 
K_{\ov T\Phi}=-\frac{H_{Z\Phi}}{2H_{ZZ}}\,,\quad
K_{\Phi\ov\Phi}=\frac{H_{Z\Phi}H_{Z\ov\Phi}}{H_{ZZ}}-H_{\Phi\ov\Phi}\,,
\end{equation}
and the determinant  \eqref{detKahler} equals  
\begin{equation}
\label{detK}
{\cal D}=-\frac{H_{\Phi\ov\Phi}}{4H_{ZZ}}\,.
\end{equation}
Putting altogether we find that the line element \eqref{genKahler} rewrites as
\begin{equation}
\label{Kahlersymmmetry}
\begin{split}
 &\text{d}s_{\text{K\"ahler}}^2=\frac{1}{4}\left(\frac{1}{V}\,\left(\text{d}\tau+\omega\right)^2+V
\left(\text{d}Z^2+\text{e}^\Psi\left(\text{d}X^2+\text{d}Y^2\right)\right)\right),\\
&V=H_{ZZ}\,,\quad\omega=H_{ZY}\text{d}X-H_{ZX}\text{d}Y\,,\quad \text{e}^\Psi=-\frac{H_{XX}+H_{YY}}{H_{ZZ}}\,,\\
&\tau=\text{Im}T\,,\quad \Phi=X+i\,Y\,,
\end{split}
\end{equation}
with\footnote{Whose compatibility relation reads:
$
V_{XX}+V_{YY}+\left(V\text{e}^\Psi\right)_{ZZ}=0\,. 
$} 
\begin{equation}
\text{d}\omega= V_X\,\text{d}Y\wedge\text{d}Z+ V_Y\,\text{d}Z\wedge\text{d}X+
\left(V\,\text{e}^\Psi\right)_Z\,\text{d}X\wedge\text{d}Y\,.
\end{equation}
These results apply to an arbitrary four-dimensional K\"ahler metric with an isometry.

\boldmath
\subsection{Hyper-K\"ahler spaces}
\unboldmath

To describe a hyper-K\"ahler metric we impose Ricci-flatness on \eqref{Kahlersymmmetry}
\begin{equation}
R_{a\ov b}=-\left(\ln{\cal D}\right)_{a\ov b}=0\,, 
\end{equation}
with general solution
\begin{equation}
{\cal D}=\text{e}^{F(T,\Phi)+\ov F(\ov T,\ov\Phi)}\,.
\end{equation}
Demanding invariance under the shift isometry of $\text{Im} T$, we find:
\begin{equation}
\label{detKsol}
{\cal D}=\text{e}^{-\alpha\left(T+\ov T\right)}|f(\Phi)|^2 \,,\quad \alpha=\text{constant}\,, 
\end{equation}
where $f(\Phi)$ can be eliminated by a holomorphic redefinition of $\Phi$, here we use $f(\Phi)=\nicefrac14$.

Employing \eqref{Legendre}, \eqref{detK} and \eqref{detKsol}, we find:
\begin{equation}
\label{Riccisol}
H_{XX}+H_{YY}+ \text{e}^{-2\alpha\,H_Z}\,H_{ZZ}=0\,.
\end{equation}

\subsubsection*{Translational isometry}

For $\alpha=0$, Eq.~\eqref{Riccisol} simplifies to the Laplace equation in Cartesian coordinates $(X,Y,Z)$
\begin{equation}
 H_{XX}+H_{YY}+H_{ZZ}=0\,,
\end{equation}
and the line element is given by 
\begin{equation}
\boxed{
\begin{split}
& \text{d}s_{\text{HK}}^2=\frac{1}{4}\left(\frac{1}{V}\,\left(\text{d}\tau+\omega\right)^2+V
\left(\text{d}X^2+\text{d}Y^2+\text{d}Z^2\right)\right),\\
&\nabla V=\nabla\times\omega\quad\Longrightarrow\quad \nabla^2V=0\,.
\end{split}
}
\end{equation}
parameterizing a hyper-K\"ahler space with a translational isometry, the Gibbons--Hawking metric \cite{Gibbons:1979zt}.

\subsubsection*{Rotational isometry}

For $\alpha\neq0$, we evaluate $\Psi$ through \eqref{Kahlersymmmetry} and \eqref{Riccisol}
\begin{equation}
\label{PsiBF}
\Psi=-2\alpha\,H_Z\,. 
\end{equation}
Next we differentiate \eqref{Riccisol} with respect to $Z$ and we get the Toda equation
\begin{equation}
\Psi_{XX}+\Psi_{YY}+\left(\text{e}^\Psi\right)_{ZZ}=0\,,
\end{equation}
and the line element is given by
\begin{equation}
\boxed{
\begin{split}
 &\text{d}s_{\text{HK}}^2=\frac{1}{4}\left(\frac{1}{V}\,\left(\text{d}\tau+\omega\right)^2+V
\left(\text{d}Z^2+\text{e}^\Psi\left(\text{d}X^2+\text{d}Y^2\right)\right)\right),\\
&V=-\frac{1}{2\alpha}\,\Psi_Z\,,\quad \omega=-\frac{1}{2\alpha}\,\Psi_Y\text{d}X+\frac{1}{2\alpha}\,\Psi_X\text{d}Y\,,\\
&\Psi_{XX}+\Psi_{YY}+\left(\text{e}^\Psi\right)_{ZZ}=0\,.
\end{split}
}
\end{equation}
parameterizing a hyper-K\"ahler space with a rotational isometry, the Boyer--Finley metric \cite{Boyer}.

\boldmath
\subsection{Scalar-flat spaces}
\unboldmath

To describe a scalar-flat space we impose vanishing scalar curvature on \eqref{Kahlersymmmetry}\footnote{So it 
is Weyl anti-self-dual, according to footnote \ref{Itoh}.}
\begin{equation}
\label{scalarflat}
 R=2g^{a\ov b}\,R_{a\ov b}=-2g^{a\ov b}\,\left(\ln{\cal D}\right)_{a\ov b}=0\,,
\end{equation}
whose general solution reads
\begin{equation}
\label{detKsol1}
{\cal D}=\text{e}^{-\alpha(T+\ov T)+\Sigma_Z}|f(\Phi)|^2 \,,\quad \alpha=\text{constant}\,,
\end{equation}
where $\Sigma_Z$ is a solution of $\nabla^2_{4\text{d}}\Sigma_Z=0$ and we use $f(\Phi)=\nicefrac14$.

Utilizing \eqref{Legendre}, \eqref{detK} and \eqref{detKsol1}, we find:
\begin{equation}
\label{Riccisol1}
H_{XX}+H_{YY}+ \text{e}^{-2\alpha\,H_Z+\Sigma_Z}\,H_{ZZ}=0\,.
\end{equation}

Using the above we obtain $\Psi$ through \eqref{Kahlersymmmetry}
\begin{equation}
\label{PsiLeBrun}
\Psi=-2\alpha\,H_Z+\Sigma_Z\,. 
\end{equation}
Next we differentiate \eqref{Riccisol1} with respect to $Z$ and we get the Toda equation
\begin{equation}
\Psi_{XX}+\Psi_{YY}+\left(\text{e}^\Psi\right)_{ZZ}=\frac14\,V\,\text{e}^\Psi\,\nabla^2_{4\text{d}}\Sigma_Z=0\,,
\end{equation}
and the line element is given by
\begin{equation}
\boxed{
\begin{split}
 &\text{d}s_{\text{LeBrun}}^2=\frac{1}{4}\left(\frac{1}{V}\,\left(\text{d}\tau+\omega\right)^2+V
\left(\text{d}Z^2+\text{e}^\Psi\left(\text{d}X^2+\text{d}Y^2\right)\right)\right),\\
&\text{d}\omega= V_X\,\text{d}Y\wedge\text{d}Z+ V_Y\,\text{d}Z\wedge\text{d}X+
\left(V\,\text{e}^\Psi\right)_Z\text{d}X\wedge\text{d}Y\,,\\
&
\Psi_{XX}+\Psi_{YY}+\left(\text{e}^\Psi\right)_{ZZ}=0\,.
\end{split}
}
\end{equation}
parameterizing a scalar-flat K\"ahler space with an isometry, the LeBrun metric \cite{LeBrun}.

\boldmath
\subsection{Einstein spaces}
\unboldmath

To describe an Einstein metric we impose on \eqref{Kahlersymmmetry}
\begin{equation}
R_{a\ov b}=-\left(\ln{\cal D}\right)_{a\ov b}=\Lambda\,K_{a\ov b}\,, 
\end{equation}
whose general solution reads
\begin{equation}
\label{detKsol2}
{\cal D}=\text{e}^{-\alpha\left(T+\ov T\right)-\Lambda\,K}|f(\Phi)|^2 \,,\quad \alpha=\text{constant}\,,
\end{equation}
where we use $f(\Phi)=\nicefrac14$.

Employing  \eqref{Legendre}, \eqref{detK} and \eqref{detKsol2}, we find:
\begin{equation}
\label{Riccisol2}
H_{XX}+H_{YY}+ \text{e}^{-2\alpha\,H_Z-\Lambda\,K}\,H_{ZZ}=0\,,
\end{equation}
Using the above we derive $\Psi$ through \eqref{Kahlersymmmetry}
\begin{equation}
\label{PsiPoon}
\Psi=-2\alpha\,H_Z-\Lambda\,K\,. 
\end{equation}
Using the latter and Eqs.~\eqref{Legendre0}, \eqref{Legendre}, \eqref{Kahlersymmmetry}, we find
\begin{equation}
\label{VPoon}
V=H_{ZZ}=-\frac{\Psi_Z}{\Lambda\left(Z+c\right)+2\alpha}\,.
\end{equation}
Then we differentiate \eqref{Riccisol2} with respect to $Z$ 
\begin{equation}
\label{TodaPoon}
\Psi_{XX}+\Psi_{YY}+\left(\text{e}^\Psi\right)_{ZZ}=
-\frac14\,\Lambda\,V\,\text{e}^\Psi\,\nabla^2_{4\text{d}}K=-2\Lambda\,V\,\text{e}^\Psi\,,
\end{equation}
where we have used:
\begin{equation}
\nabla^2_{4\text{d}}K=2g^{a\ov b}\nabla_a\nabla_{\ov b} K=2g^{a\ov b}\, K_{a\ov b}
=2\,g^{a\ov b}\,g_{a\ov b}=8\,,\quad \Gamma_{a\ov b}{}^c=\Gamma_{a\ov b}{}^{\ov c}=0\,.
\end{equation}
Plugging \eqref{VPoon} into \eqref{TodaPoon} we find the Pedersen--Poon equation
\begin{equation}
\Psi_{XX}+\Psi_{YY}+\left(\text{e}^\Psi\right)_{ZZ}=
\frac{2\Lambda\,\Psi_Z\,\text{e}^\Psi}{\Lambda\left(Z+c\right)+2\alpha}\,,
\end{equation}
and the line element is given by 
\begin{equation}
\boxed{
\begin{split}
 &\text{d}s_{\text{PP}}^2=\frac{1}{4}\left(\frac{1}{V}\,\left(\text{d}\tau+\omega\right)^2+V
\left(\text{d}Z^2+\text{e}^\Psi\left(\text{d}X^2+\text{d}Y^2\right)\right)\right),\\
&\text{d}\omega= V_X\,\text{d}Y\wedge\text{d}Z+ V_Y\,\text{d}Z\wedge\text{d}X+
\left(V\,\text{e}^\Psi\right)_Z\,\text{d}X\wedge\text{d}Y\,,\\
&V=-\frac{\Psi_Z}{\Lambda\left(Z+c\right)+2\alpha}\,,\quad
\Psi_{XX}+\Psi_{YY}+\left(\text{e}^\Psi\right)_{ZZ}=
\frac{2\Lambda\,\Psi_Z\,\text{e}^\Psi}{\Lambda\left(Z+c\right)+2\alpha}\,.
\end{split}
}
\end{equation}
parameterizing a K\"ahler--Einstein space with an isometry, the Pedersen--Poon metric \cite{Poon}.


\end{document}